\documentclass[12pt,a4paper]{article}
\usepackage{amsmath,bbm,longtable,paralist}
\usepackage{caption}
\usepackage[top=1.5in, bottom=1.5in, left=1.1in, right=1.1in]{geometry}

\DeclareCaptionStyle{italic}[justification=centering]{labelfont={bf},textfont={it},labelsep=colon}
\usepackage{graphicx,psfrag,epsf}
\usepackage{enumerate}
\usepackage{natbib}
\usepackage{url}
\usepackage{bbm, dsfont}
\usepackage{comment}
\usepackage{geometry}
\usepackage{float}
\usepackage[dvipsnames, table]{xcolor}
\usepackage{booktabs, subfig, bm, paralist,mathpazo,tikz,todonotes,longtable,microtype,dsfont,rotating} 
\usepackage[pdftex,colorlinks=true]{hyperref}
\definecolor{darkblue}{rgb}{0,0,.6}
\hypersetup{citecolor=darkblue,linkcolor=darkblue,urlcolor=darkblue}

\newcommand{\blind}{0}

\graphicspath{{plots/}}
\DeclareMathOperator*{\argmin}{\arg\!\min}
\newsavebox\CBox
\def\textBF#1{\sbox\CBox{#1}\resizebox{\wd\CBox}{\ht\CBox}{\textbf{#1}}}

\definecolor{a0}{rgb}{0.0, 0.5, 0.0}
\definecolor{bistre}{rgb}{0.24, 0.17, 0.12}
\definecolor{amethyst}{rgb}{0.6, 0.4, 0.8}
\definecolor{blue-violet}{rgb}{0.54, 0.17, 0.89}
\definecolor{Rcolor}{RGB}{150,160,190}
\definecolor{blush}{rgb}{0.87, 0.36, 0.51}
\definecolor{brightturquoise}{rgb}{0.03, 0.91, 0.87}
\definecolor{burntorange}{rgb}{0.8, 0.33, 0.0}

\usepackage{orcidlink}
\date{}
\AtBeginDocument{}
\newcommand{\Rlogo}{\protect\includegraphics[height=1.8ex,keepaspectratio]{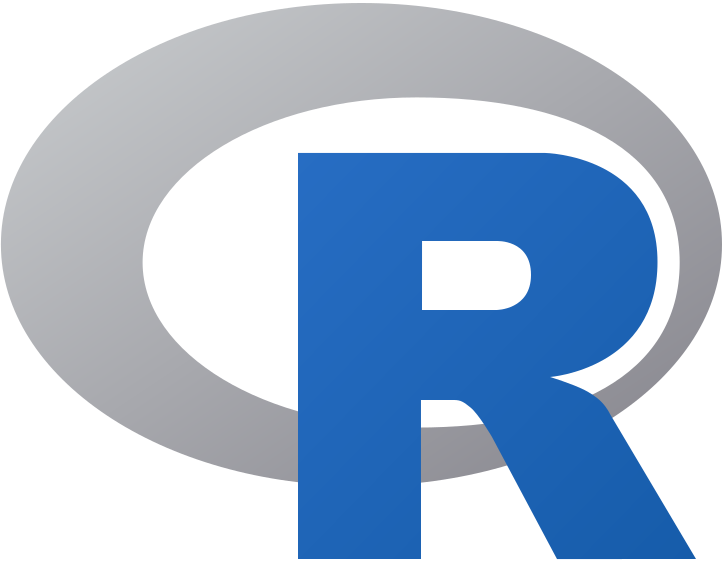}}
\begin{document}

\def\spacingset#1{\renewcommand{\baselinestretch}{#1}\small\normalsize} \spacingset{1}

\if0\blind
{
  \title{\bf Constructing prediction intervals for the age distribution of deaths}
  \author{\normalsize Han Lin Shang \orcidlink{0000-0003-1769-6430} \footnote{Postal address: Department of Actuarial Studies and Business Analytics, Macquarie University, Sydney, NSW 2109, Australia; Telephone: +61(2) 9850 4689; Email: hanlin.shang@mq.edu.au}\\
    \normalsize Department of Actuarial Studies and Business Analytics \\
    \normalsize Macquarie University \\
    \\
\normalsize    Steven Haberman \orcidlink{0000-0003-2269-9759}\\
\normalsize Bayes Business School \\
\normalsize City St George's, University of London
}
  \maketitle
} \fi

\if1\blind
{
\title{\bf Constructing prediction intervals for the age distribution of deaths}
  \maketitle
} \fi

\bigskip

\begin{abstract}
We introduce a model-agnostic procedure to construct prediction intervals for the age distribution of deaths. The age distribution of deaths is an example of constrained data, which are nonnegative and have a constrained integral. A centered log-ratio transformation and a cumulative distribution function transformation are used to remove the two constraints, where the latter transformation can also handle the presence of zero counts. Our general procedure divides data samples into training, validation, and testing sets. Within the validation set, we can select an optimal tuning parameter by calibrating the empirical coverage probabilities to be close to their nominal ones. With the selected optimal tuning parameter, we then construct the pointwise prediction intervals using the same models for the holdout data in the testing set. Using Japanese age- and sex-specific life-table death counts, we assess and evaluate the interval forecast accuracy with a suite of functional time-series models.

\vspace{.1in}
\end{abstract}

\newpage
\spacingset{1.6}

\section{Introduction}\label{sec:intro}

Actuaries and demographers have long been interested in developing statistical techniques to model and forecast mortality for annuity pricing and government planning. In the literature on human mortality, three functions are widely studied: mortality rate, survival function, and life-table death counts (representing age distribution of deaths). Although these three functions are complementary \citep[see, e.g.,][]{PHG01, DMW09}, they differ by the number of constraints. The mortality rate is between 0 and 1; the survival function is also between 0 and 1 and exhibits monotonicity over a certain age group; and the life-table death counts are non-negative and sum up to a radix, commonly $10^5$.

Most of the literature has focused on the development of novel approaches for modeling and forecasting age-specific logarithmic mortality rates \citep[see, e.g.,][for comprehensive reviews]{Booth06, BT08, BCB23}. Instead of modeling central mortality rates, we consider modeling life-table deaths as an example of a probability density function \citep[see, e.g.,][]{BKC20}. Observed over a period, we could visualize, model, and forecast a redistribution of the life-table deaths, where deaths at younger ages are shifted gradually toward older ages due to longevity. In addition to providing an informative description of the mortality experience of a population, life-table deaths provide readily available information on central longevity indicators \citep[see, e.g.,][]{CRT+05, CanudasRomo10} and lifespan variability \citep[see, e.g.,][]{Robine01, VZV11, VC13, AV18, AVB+20}. 

To model the age distribution of deaths, we resort to an extrinsic approach via transformation. In demography, \cite{BCO+17} and \cite{BSO+18} apply the centered log ratio (CLR) transformation to obtain unconstrained data, which can then be modeled through principal component analysis. In actuarial science, \cite{SH20} and \cite{SHX22} used the forecasted life-table death counts to calculate estimated fixed-term annuity prices. In statistics, \cite{SM22} apply the CLR transformation to model cause-specific mortality data, \cite{Delicado11} apply the CLR transformation to analyze density functions over space, and \cite{KMP+19} model and forecast financial time series of density functions.

An issue with the CLR transformation is the presence of zero counts. Some ad-hoc ways of handling zero counts exist, including adding or subtracting a small constant \citep[see, e.g.,][]{MBP13, FFM00}. Recently, \cite{SH25} introduced a cumulative distribution function (CDF) transformation with the advantage of monotonicity. We first normalize the age distribution of death so that the radix is one, akin to the probability density function (PDF), and then convert the PDF to a CDF. The inverse of CDF is quantile, which is a key quantity in the Wasserstein distance to measure the discrepancy between two distributions \citep[see, e.g.,][]{DM22}. With a time series of CDFs, we model its pattern via a logistic transformation. Within this unconstrained space, we apply a suite of functional time-series forecasting methods to obtain the $h$-step-ahead curve prediction for a chosen forecast horizon $h$. By taking the inverse logistic transformation, the $h$-step-ahead forecast life-table death counts are obtained after first-order differencing and renormalized to the original scale.

The current literature lacks guidance on the construction of prediction intervals for the age distribution of deaths. We aim to present a general procedure that works for time-series forecasting models in Section~\ref{sec:3}. The general procedure divides the data samples into training, validation, and testing sets. The validation set allows us to tune an optimal parameter that adjusts the prediction intervals so that the empirical coverage probability is close to its nominal one. With the selected optimal parameter, we construct the pointwise prediction intervals for the data in the holdout set. Assuming that the data in the validation and testing sets do not differ much, our construction can achieve satisfactory coverage. Using age- and sex-specific life-table death counts in Japan in Section~\ref{sec:2}, we study the interval forecast accuracy of several functional time-series methods in Section~\ref{sec:4}. The conclusion is presented in Section~\ref{sec:5}, along with some ideas on how the methodology can be further extended.

\section{Period life-table death counts}\label{sec:2}

In many developed countries, such as Japan, increases in longevity risk and an aging population have led to concerns about the sustainability of government pension, health and age care systems. Japan has one of the highest average life expectancies in the world, with extreme longevity in Okinawa prefecture \citep{Coulmas07}.

Our chosen mortality instrument is the life-table death counts, where the life-table radix is fixed at $100,000$ at age zero while the remaining number of people alive is 0 in the last age group 110+ for each year. There are 111 ages, which are $0, 1, \dots, 109, 110+$. Due to rounding, there are potentially zero counts for people aged 110+ at some years. To overcome this problem, we work with the probability of dying (i.e., $q_x$) and the radix of the life table to recalculate our estimated death counts (up to six decimal places). In doing so, we obtain more precise death counts than those reported in \cite{JMD25}. 

\begin{figure}[!htb]
\centering
\includegraphics[width=8.7cm]{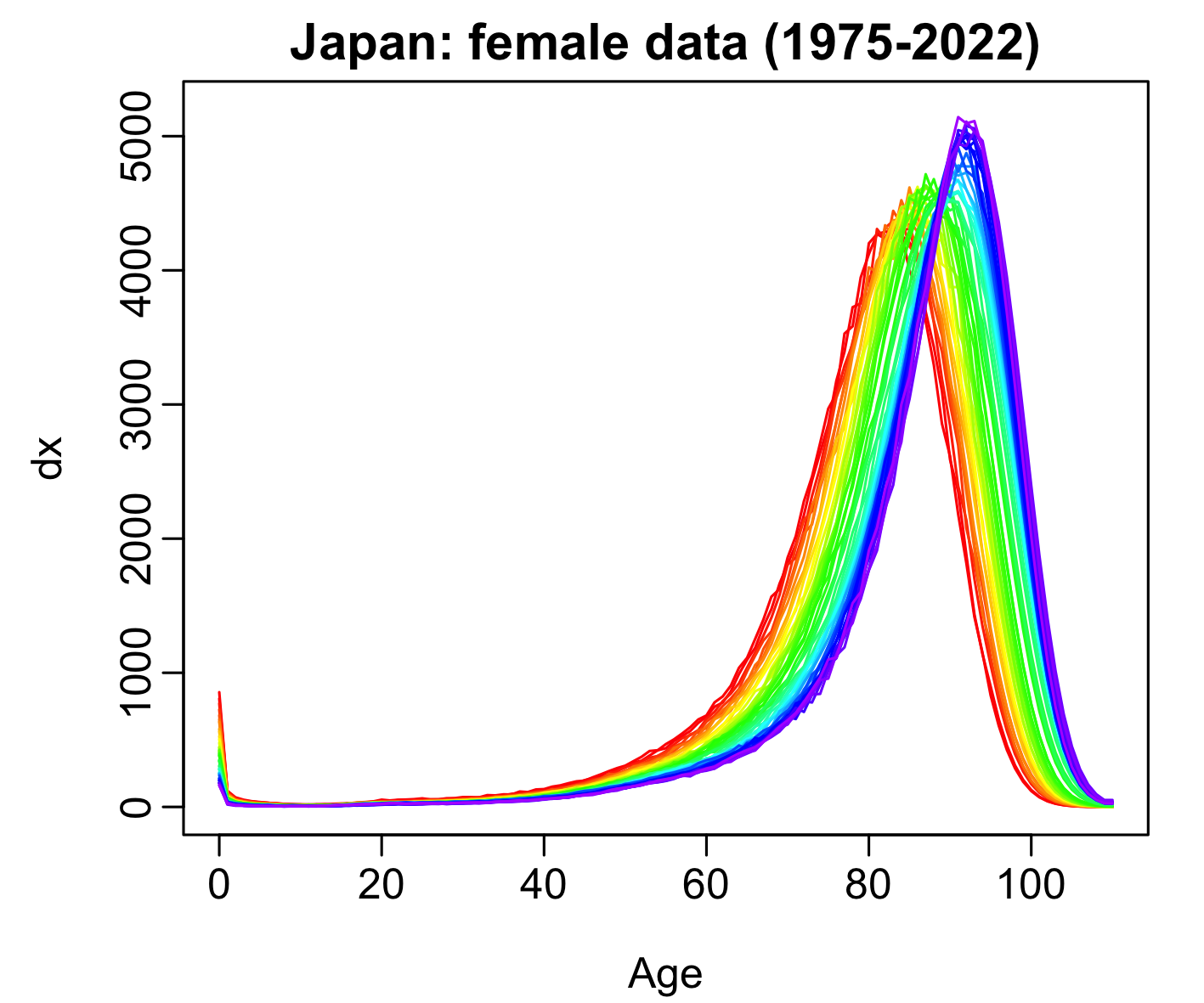}
\quad
\includegraphics[width=8.7cm]{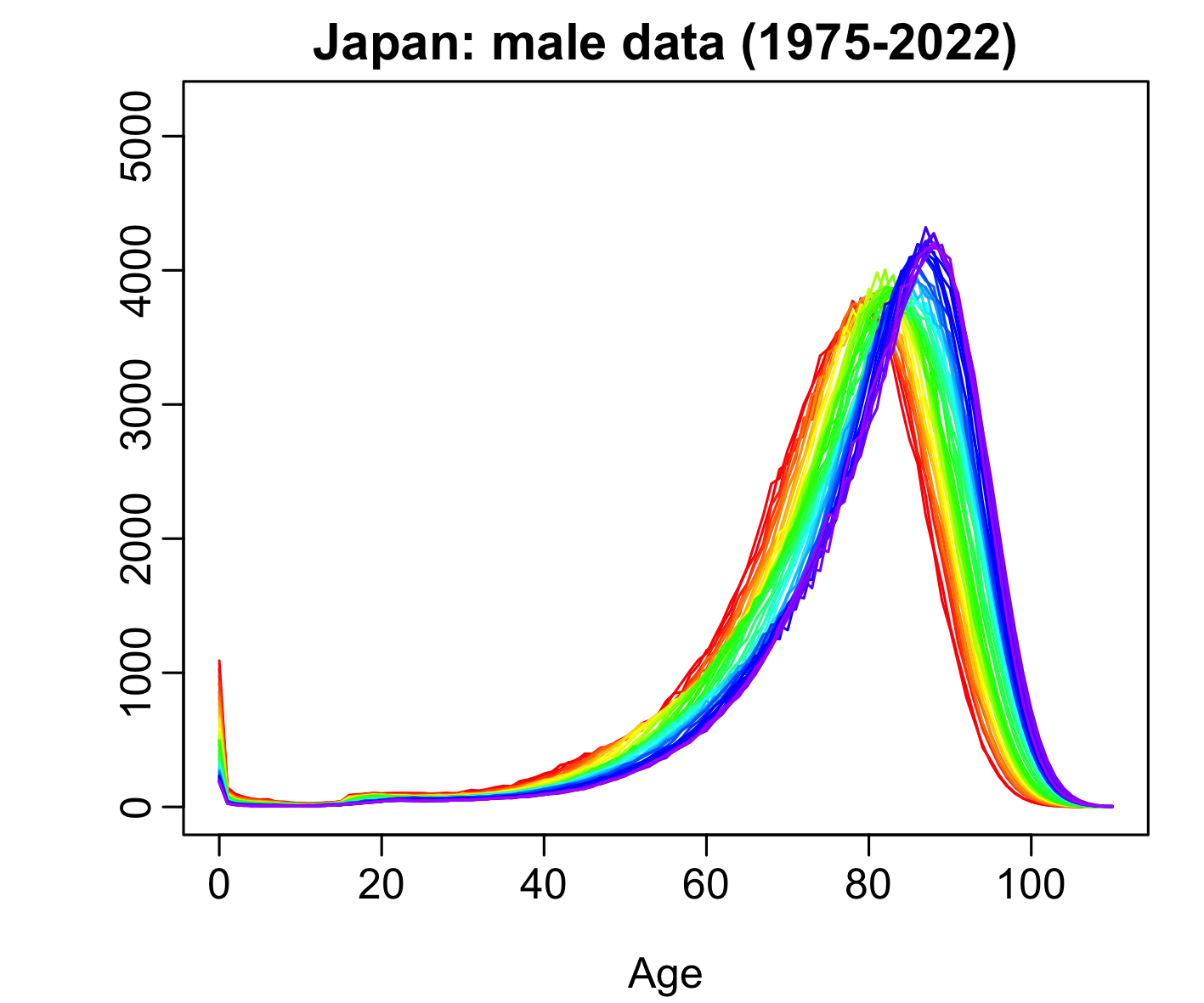}
\caption{\small Rainbow plots of the age distribution of deaths from 1975 to 2022 in a single-year group. The life-table radix is 100,000 for each year. The life-table death counts in the oldest years are shown in red, while the most recent years are in violet. Curves are ordered chronologically by the rainbow colors.}\label{fig:1}
\end{figure}

In Figure~\ref{fig:1}, we present Japanese age- and sex-specific life-table death counts from 1975 to 2022, obtained from \cite{JMD25}. We used data from the period after the First and Second World Wars to obtain a more stable parameter estimate from the historical data. The beginning year 1975 was chosen to be the same as its subnational data \citep[see also][]{SH25}.

Figure~\ref{fig:1} demonstrates a decreasing trend in infant death counts and a typical negatively skewed distribution for life-table death counts, where the peak shifts to higher ages for both sexes. This shift is due to the risk of longevity, which concerns insurers and pension funds in transferring and managing the risks of annuity products \citep[see][for a discussion]{DDG07}. By modeling the period life-table death counts, we can understand a redistribution of life-table death counts, where deaths at younger ages gradually shift towards older ages.

Since the exposure-to-risk can be difficult to estimate accurately due to migration, under-reporting, or late registration \citep[see, e.g.,][]{CBD+16}, we choose to work with life-table death counts instead of mortality rates. Life-table death counts are derived from the probability of dying and bypass the need for direct exposure estimation, and they represent the number of deaths in the implied stationary population and lead to the corresponding probability density function of the age distribution of deaths. Due to non-negativity and summability constraints, we can study the age distribution of deaths for all available ages. 

\section{Construction of prediction intervals for density-valued objects}\label{sec:3}

Denote age-specific life-table death counts as $d_{t}^s(u)$, where $t$ denotes a year, $s$ denotes female or male data, and $u$ represents an age. For each year $t$, the life-table death counts sum to a radix~$10^5$. In Sections~\ref{sec:3.1} and~\ref{sec:3.2}, we consider two transformation methods to remove constraints in the life-table death counts. For modeling the unconstrained data within each transformation, we consider three functional time-series forecasting models in Section~\ref{sec:3.3}.

\subsection{Centered log-ratio transformation}\label{sec:3.1}

By treating age as a continuous variable, the CLR transformation can be written as
\begin{equation*}
\text{CLR}[d_t^s(u)] := G_t^s(u) = \ln d_t^s(u) - \frac{1}{\eta}\int_{u}\ln d_t^s(u) du,
\end{equation*}
where $\eta$ denotes the length of the age interval and $\frac{1}{\eta}\int_{u}\ln d_t^s(u) du$ is the geometric mean. With a time series of functions $[G_1^s(u),\dots, G_n^s(u)]$, we apply the univariate, multivariate, and multilevel functional time-series models to obtain $h$-step-ahead forecasts $\widehat{G}_{n+h|n}^s(u)$. A brief description of these time-series models is given in Section~\ref{sec:3.3}. Via the inverse CLR transformation, the forecast life-table death counts can be expressed as
\begin{equation*}
\widehat{d}_{n+h|n}^s(u) = \frac{\exp^{\widehat{G}_{n+h|n}(u)}}{\int_{u} \exp^{\widehat{G}_{n+h|n}(u)}du} \times 10^5.
\end{equation*}

\subsection{Cumulative distribution function transformation}\label{sec:3.2}

By normalizing the life-table radix from $10^5$ to one, the first transformation computes the empirical CDF via cumulative sum,
\begin{equation*}
D_{t}^s(x) = \sum^x_{u=1}d_{t}^s(u), \quad x=1,\dots,111.
\end{equation*}
Since $D_t^{s}(111)=1$, we apply the logistic transformation to the first 110 ages, 
\begin{equation*}
L_t^s(y) = \ln\Big[\frac{D_{t}^s(y)}{1-D_{t}^s(y)}\Big],\quad y = 1,\dots, 110,
\end{equation*}
where $\ln(\cdot)$ denotes the natural logarithm. With a time series of functions $[L_1^s(y),\dots, L_{n}^s(y)]$, we obtain $h$-step-ahead forecasts, denoted by $\widehat{L}_{n+h|n}^s(y)$, via the univariate, multivariate, and multilevel functional time-series methods. 

Taking the inverse logit transformation, we obtain
\begin{equation*}
\widehat{D}_{n+h|n}^s(y) = \frac{\exp^{\widehat{L}_{n+h|n}^s(y)}}{1+\exp^{\widehat{L}_{n+h|n}^s(y)}}.
\end{equation*}
By including a column of ones $\bm{1}$, we obtain $\widehat{D}^s_{n+h|n}(x)=[\widehat{D}^s_{n+h|n}(y),\bm{1}]$. By taking the first-order differencing, we obtain
\begin{align*}
\widehat{d}_{n+h|n}^s(z) &= \Delta  \widehat{D}^s_{n+h|n}(z)  \\
&= \widehat{D}^s_{n+h|n}(z) - \widehat{D}^s_{n+h|n}(z-1),\quad z=2,\dots,111,
\end{align*}
where $\Delta$ represents the first-order differencing, and $\widehat{d}^s_{n+h|n}(1)=\widehat{D}^s_{n+h|n}(1)$. Given the life-table radix of $10^5$, we renormalize the forecasts to their original scale: $\widehat{d}_{n+h|n}^s(u) = \widehat{d}_{n+h|n}^s(z)\times 10^5$.

\subsection{A suite of functional time-series forecasting methods}\label{sec:3.3}

The unconstrained data are assumed to be elements of the Hilbert space equipped with the inner product. We model the unconstrained data, $G_t^s(u)$ in the CLR transformation or $L_t^s(u)$ in the CDF transformation. For illustration, we demonstrate our idea with $G_t^s(u)$, which can be expressed via the Karhunen-Lo\`{e}ve expansion as
\begin{equation}
G_t^s(u) = \sum_{k=1}^{K_s}\eta_{t,k}^s\psi_{k}^s(u)+\epsilon_{t}^s(u), \tag{UFTS}
\end{equation}
where $\psi_k^s(u)$ denotes the $k$\textsuperscript{th} functional principal component for age $u$ and sex $s$, $\eta_{t,k}^s = \langle G_t^s(u), \psi_k^s(u)\rangle$ is the estimated principal component score at time $t$ and $\langle \cdot,\cdot\rangle$ denotes the $L^2$ inner product, $\epsilon_t^s(u)$ denotes the residual function for age $u$ and sex $s$ in year $t$, and $K_s$ denotes the number of functional principal components. We consider an eigenvalue ratio (EVR) criterion of \cite{LRS20} to select the number of $K_s$, which is the integer that minimizes the ratio of two adjacent empirical eigenvalues given by 
\begin{equation}
K_s = \argmin_{1\leq \kappa\leq (n-1)}\left\{\frac{\widehat{\lambda}^s_{\kappa+1}}{\widehat{\lambda}^s_{\kappa}}\times \mathds{1}(\frac{\widehat{\lambda}^s_{\kappa+1}}{\widehat{\lambda}^s_{\kappa}}\geq \delta)+\mathds{1}(\frac{\widehat{\lambda}^s_{\kappa+1}}{\widehat{\lambda}^s_{\kappa}}<\delta) \right\},\label{eq:EVR}
\end{equation}
where $\widehat{\lambda}^s_{\kappa}$ is the $\kappa$\textsuperscript{th} estimated eigenvalue, $\delta$ is a prespecific small positive number, set as $\delta=1/\ln(\max\{\widehat{\lambda}^s_1,n\})$, and $\mathds{1}(\cdot)$ denotes the binary indicator function. For comparison, we also consider $K_s = 6$ used in \cite{HBY13}.

We also consider a multivariate functional time-series method to jointly model and forecast the female and male series that could be correlated. Let $G_t^F(u)$ and $G_t^M(u)$ represent unconstrained female and male data. By stacking both series in a vector, we compute their joint covariance function. Via Karhunen-Lo\`{e}ve expansion, a realization of both series can be approximated by
\begin{equation}
\bm{G}_t(u) = \bm{\theta}(u) + \bm{\Phi}(u)\bm{\beta}_{t}^{\top}, \tag{MFTS}
\end{equation}
where $\bm{G}_t(u) = [G_t^{\text{F}}(u), G_t^{\text{M}}(u)]^{\top}$; $\bm{\theta}(u) = [\theta^{\text{F}}(u),\theta^{\text{M}}(u)]^{\top}$ denotes the mean functions for the female and male series, respectively; $\bm{\Phi}(u)$ is a $(2\times (K\times 2))$ matrix, where the off-diagonal elements capture the correlation between the estimated principal components; $\bm{\beta}_t=[\bm{\beta}^{\text{F}}_t, \bm{\beta}^{\text{M}}_t]$ and $\bm{\beta}_{t}^{\text{F}} = [\beta^{\text{F}}_{t, 1},\dots,\beta_{t, K}^{\text{F}}]$ denotes the estimated principal component scores.

The multilevel functional time-series method extracts a common pattern shared by female and male series $R_t(u)$ and a series-specific residual pattern $U_t^s(u)$. Via functional principal component analysis, the common and residual patterns are modeled by projecting the data onto the eigenfunctions of the covariance functions of aggregated and series-specific curves, respectively. For $t=1,2,\dots,n$, a realization can be approximated by
\begin{equation}
G_{t}^{s}(u) = \mu^s(u) + R_t(u) + U_t^s(u). \tag{MLFTS}
\end{equation}
With a finite sample, we estimate 
\begin{align*}
\widehat{\mu}^s(u) &= \frac{1}{n}\sum^n_{t=1}G_t^s(u) \\
R_t(u) &\approx \sum_{k=1}^K\beta_{t,k}\phi_k(u) \\
U_t^s(u) &\approx \sum_{\ell=1}^{V} \gamma_{t,\ell}\psi_{\ell}(u),
\end{align*}
where $K$ and $V$ represent the number of functional principal components retained. These components can be determined by the EVR criterion in~\eqref{eq:EVR} or set to $K=V=6$.

\subsection{Construction of prediction intervals}\label{sec:3.4}

We equally divide the data sample consisting of 48 years from 1975 to 2022 into training, validation, and testing sets, each consisting of 16 years. Using the data in the training sample, we implement an expanding window forecast scheme to obtain the $h$-step-ahead density forecasts in the validation set for $h=1, 2, \dots, 15$. The expanding window scheme allows one to assess how a forecasting method performs on short and medium horizons. We have different numbers of curves in the validation set for each forecast horizon. For example, when $h=1$, we have 16 years to evaluate the forecast errors; when $h=15$, we have two years to evaluate the residual functions between the samples in the validation set and their forecasts, and compute their functional standard deviation. Note that we need at least two years of data to compute the functional standard deviation. Forecast errors are denoted by $\widehat{\varepsilon}_m(u) = d_m^s(u) - \widehat{d}_m^s(u)$, for $m=1, 2, \dots, M$, and $M$ denotes the number of years of residual functions. 

Let us define $\gamma(u) = \text{sd}[\widehat{\varepsilon}_m(u)]$. For a level of significance $\alpha$, our aim is to determine ($\underline{\xi}_{\alpha}, \overline{\xi}_{\alpha}$) such that $\alpha\times 100\%$ of the residuals satisfy
\begin{equation*}
-\underline{\xi}_{\alpha}\gamma(u)\leq \widehat{\varepsilon}_m(u)\leq \overline{\xi}_{\alpha}\gamma(u).
\end{equation*} 
($\overline{\xi}_{\alpha}, \underline{\xi}_{\alpha}$) are the tuning parameters mentioned in the abstract and Section~\ref{sec:intro}. Typically, the constants $\overline{\xi}_{\alpha}$ and $\underline{\xi}_{\alpha}$ are chosen equal. By the law of large numbers, one may achieve
\begin{equation*}
\text{Pr}[-\xi_{\alpha}\gamma(u)\leq d^s_{n+h}(u)-\widehat{d}^s_{n+h|n}(u) \leq \xi_{\alpha}\gamma(u)]\approx\frac{1}{M}\sum^M_{m=1}\mathds{1}[-\xi_{\alpha}\gamma(u)\leq \widehat{\varepsilon}_m(u)\leq \xi_{\alpha}\gamma(u)].
\end{equation*}

To determine the optimal $\xi_{\alpha}$, the samples in the validation set are used to calibrate the prediction intervals so that the empirical coverage probabilities are close to their nominal coverage probabilities. As an output of this calibration, we obtain an optimal tuning parameter based on the coverage probability difference in Section~\ref{sec:4.2}.

For comparison, we also consider conformal prediction intervals, which are well calibrated in a large sample size \citep{DDR24}. The conformal prediction introduced by \cite{SV08} is a popular methodology in machine learning and is used to construct probabilistic forecasts calibrated on out-of-sample errors. Since its introduction in \cite{GVV98}, it has received increasing attention in various fields, including time series forecasting \citep{YX21, FZV23, ACT23} and climate modeling \citep{Cannon18, QC21}. The conformal prediction is model-agnostic and presents a distribution-free way to construct prediction sets with a finite-sample coverage guarantee. From the absolute value of $\widehat{\varepsilon}_m(u)$, we calculate its $100(1-\alpha)\%$ quantile for a level of significance $\alpha$, denoted by $q_{\alpha}(u)$. The prediction interval can be obtained as 
\begin{equation*}
\left[\widehat{d}^s_{n+h|n}(u)-q_{\alpha}(u), \widehat{d}^s_{n+h|n}(u)+q_{\alpha}(u)\right],
\end{equation*}
where $\widehat{d}^s_{n+h|n}(u)$ denotes the $h$-step-ahead point forecasts for the data in the test set. 

We consider the simplest conformity score by taking the quantiles from the absolute residuals. We acknowledge that other conformity scores, such as the use of quantile regression, are possible to construct asymmetric prediction intervals that may lead to better performance \citep[see, e.g.,][]{RPC19, CWZ21}. Two limitations are commonly associated with the split conformal prediction: First, it works well for identically distributed data under the assumption of exchangeability. For time series data, the empirical coverage deteriorates as the forecast horizon increases. Second, it requires a large sample size for the validation and testing sets to achieve superior calibration.

\section{Evaluation of interval forecast accuracy}\label{sec:4}

\subsection{Expanding-window forecast scheme}\label{sec:4.1}

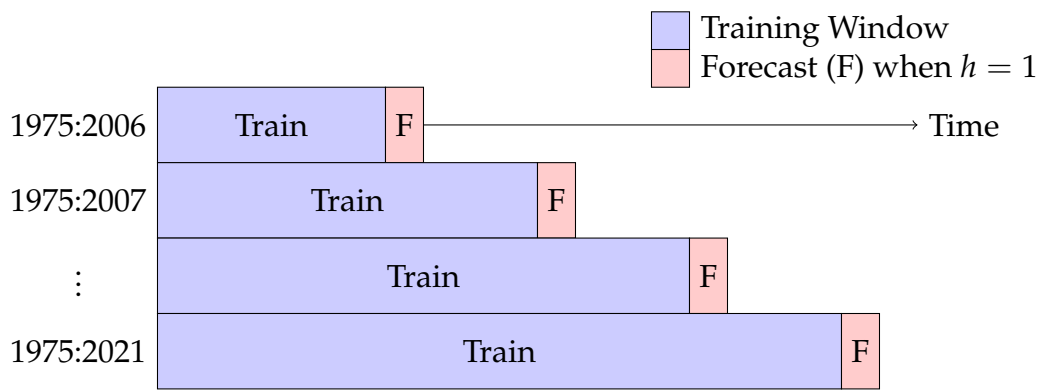
\begin{figure}[!htb]
\begin{center}
\begin{tikzpicture}
\draw[->] (0,0) -- (10,0) node[right] {Time};
    
\draw[fill=blue!20] (0,-0.5) rectangle (3,0.5) node[midway] {Train};
\draw[fill=red!20] (3,-0.5) rectangle (3.5,0.5) node[midway] {F};
    
\draw[fill=blue!20] (0,-1.5) rectangle (5,-0.5) node[midway] {Train};
\draw[fill=red!20] (5,-1.5) rectangle (5.5,-0.5) node[midway] {F};
    
\draw[fill=blue!20] (0,-2.5) rectangle (7,-1.5) node[midway] {Train};
\draw[fill=red!20] (7,-2.5) rectangle (7.5,-1.5) node[midway] {F};
    
\draw[fill=blue!20] (0,-3.5) rectangle (9,-2.5) node[midway] {Train};
\draw[fill=red!20] (9,-3.5) rectangle (9.5,-2.5) node[midway] {F};
    
\node[left] at (0,0) {1975:2006};
\node[left] at (0,-1) {1975:2007};
\node[left] at (0,-2) {\hspace{-0.8in}{$\vdots$}};
\node[left] at (0,-3) {1975:2021};
    
\draw[fill=blue!20] (6.5,1) rectangle (7,1.5);
\node[right] at (7,1.25) {Training Window};
\draw[fill=red!20] (6.5,0.5) rectangle (7,1);
\node[right] at (7,0.75) {Forecast (F) when $h=1$};
\end{tikzpicture}
\end{center}
\caption{A diagram of the expanding-window forecast scheme.}\label{fig:expanding}
\end{figure}

An expanding window analysis of a time-series model is commonly used to assess model and parameter stability over time. With the samples in the test set, we evaluate and assess the accuracy of the interval forecast. Using the first 32 years from 1975 to 2006, we can produce one- to 16-step-ahead forecasts. Through an expanding window scheme, we estimate the parameters in the time-series forecasting models using the first 33 observations from 1975 to 2007. Forecasts from the estimated model are produced for one- to 15-step-ahead forecasts. We iterate this process by increasing the sample size by one year until we reach the end of the data period in 2022. This iteration process produces 16 one-step-ahead forecasts, 15 two-step-ahead forecasts, $\dots$, and one 16-step-ahead forecast. In Figure~\ref{fig:expanding}, we show a diagram of the expanding window forecast scheme for the forecast horizon $h=1$, although we also consider other forecast horizons from $h=2$ to 15.

\subsection{Interval forecast errors}\label{sec:4.2}

To evaluate interval forecast accuracy, we consider the coverage probability difference (CPD) between the empirical coverage probability (ECP) and nominal coverage probability, as well as the mean interval score of \cite{GR07}. For each year in the forecast period, the $h$-step-ahead prediction intervals are calculated at the $100(1-\alpha)\%$ nominal coverage probability. We consider the common case of the symmetric $100(1-\alpha)\%$ prediction intervals, with lower and upper bounds that are quantiles at $\alpha/2$ and $1-\alpha/2$, denoted by $\widehat{d}_{n+\xi}^{s, \text{lb}}(u)$ and $\widehat{d}_{n+\xi}^{s, \text{ub}}(u)$. The ECP and CPD are defined as
\begin{align*}
\text{ECP}_h &=  \frac{1}{111\times (16-h)}\times \sum_{\xi=h}^{16}\sum_{u=1}^{111}\mathds{1}\left\{\widehat{d}_{\nu+\xi}^{s, \text{lb}}(u)\leq d_{\nu+\xi}(u) \leq  \widehat{d}_{\nu+\xi}^{s, \text{ub}}(u)\right\}, \\
\text{CPD}_h &= \frac{1}{111\times (16-h)}\times \sum_{\xi=h}^{16}\sum_{u=1}^{111}\left[\mathds{1}\{d_{\nu+\xi}(u)>\widehat{d}_{\nu+\xi}^{s, \text{ub}}(u)\}+\mathds{1}\{d_{\nu+\xi}(u)<\widehat{d}_{\nu+\xi}^{s, \text{lb}}(u)\}\right],
\end{align*}
where $v$ denotes the years in the training and validation sets.

For different ages and years in the test set, the mean and median ECP are defined as
\begin{align*}
\overline{\text{ECP}} &= \frac{1}{15}\text{ECP}_h, \\
\text{M[ECP]} &= \text{median}(\text{ECP}_h).
\end{align*}
Similarly, mean and median CPD are defined as 
\begin{align*}
\overline{\text{CPD}} &= \frac{1}{15}\sum^{15}_{h=1}\text{CPD}_h, \\
\text{M[CPD]} &=\text{median}(\text{CPD}_h).
\end{align*}

As defined by \cite{GR07}, a scoring rule for the prediction intervals at age $u$ is
\begin{align*}
S_{\alpha,\xi}\left[\widehat{d}^{s,\text{lb}}_{\nu+\xi}(u), \widehat{d}^{s,\text{ub}}_{\nu+\xi}(u), d^s_{\nu+\xi}(u)\right] = \ &  \left[\widehat{d}^{s,\text{ub}}_{\nu+\xi}(u) - \widehat{d}^{s,\text{lb}}_{\nu+\xi}(u)\right] \\
&+\frac{2}{\alpha}\left[\widehat{d}^{s,\text{lb}}_{\nu+\xi}(u)-d_{n+\xi}^s(u)\right]\mathds{1}\left\{d_{\nu+\xi}^s(u) <\widehat{d}_{\nu+\xi}^{s,\text{lb}}(u)\right\} \\
&+\frac{2}{\alpha}\left[d_{\nu+\xi}^s(u) - \widehat{d}_{\nu+\xi}^{s,\text{ub}}(u)\right]\mathds{1}\left\{d_{\nu+\xi}(u) > \widehat{d}_{\nu+\xi}^{s,\text{ub}}(u)\right\},
\end{align*}
where the level of significance is customarily set to $\alpha = 0.2$ or 0.05. The interval score rewards a narrow prediction interval width if and only if $100(1-\alpha)\%$ of the holdout densities lies within the prediction interval.

For different ages and years in the test set, the mean interval score is defined by
\begin{align*}
\overline{S}_{\alpha}(h) = \frac{1}{111\times (16-h)}\times \sum^{16}_{\xi=h}\sum^{111}_{u=1}S_{\alpha,\xi}\left[\widehat{d}_{\nu+\xi}^{s,\text{lb}}(u), \widehat{d}_{\nu+\xi}^{s,\text{ub}}(u), d_{\nu+\xi}^s(u)\right].
\end{align*}
Averaging over all forecast horizons, we obtain the overall mean interval score 
\begin{align*}
\overline{S}_{\alpha} &= \frac{1}{15}\sum^{15}_{h=1}\overline{S}_{\alpha}(h), \\ 
\text{M}[S_{\alpha}] &= \text{median}[\overline{S}_{\alpha}(h)].
\end{align*}

\subsection{Interval forecast results}\label{sec:4.3}

For $h = 1,2,\dots,15$, we present the estimated values of $\xi_{\alpha}$ obtained from the univariate functional time-series model with the CDF transformation in Table~\ref{tab:1}. Regardless of the method used to select the number of principal components retained, the values of $\xi_{\alpha}$ exhibit an increasing trend as $h$ increases. This pattern highlights the increasing uncertainty associated with longer-term forecasts. When $h=15$, there exists a numerical instability issue since we have only 2 years of data samples in the validation set.
\begin{table}[!htb]
\centering
\tabcolsep 0.25in
\renewcommand*{\arraystretch}{0.63}
\caption{\small For different forecast horizons $h = 1, 2, \dots,15$, we present the estimated tuning parameter $\xi_{\alpha}$ values obtained from the univariate functional time-series model (the number of retained principal components, $K$, can be determined via the EVR criterion or set as six) with the CDF transformation.}\label{tab:1}
\begin{tabular}{@{}rrrrrrrrr@{}}
\toprule
& \multicolumn{4}{c}{$\alpha=0.2$} & \multicolumn{4}{c}{$\alpha=0.05$} \\
& \multicolumn{2}{c}{EVR}   & \multicolumn{2}{c}{$K=6$}   & \multicolumn{2}{c}{EVR}   & \multicolumn{2}{c}{$K=6$} \\
$h$ & F & M & F & M & F & M & F & M \\ 
\midrule
1 & 1.41  & 1.55  & 1.37 &  1.40 &  2.09 &  2.28  & 1.98  & 2.28  \\ 
2 & 1.42  & 1.52 & 1.39 & 1.43 &  2.09 &  2.28  & 1.98  & 2.28  \\ 
3 & 1.47  & 1.57 & 1.44 & 1.47 &  2.16 &  2.32  & 2.04  & 2.32  \\ 
4 & 1.57  & 1.66  & 1.53 & 1.59 & 2.35 &  2.45  & 2.25  & 2.45 \\ 
5 & 2.00  & 1.95  & 1.95 &1.76 &  2.84 &  2.85  & 2.78  & 2.85  \\ 
6 & 2.28  & 3.26  & 2.22 & 2.99 & 3.40 &  4.50 & 3.35  & 4.50  \\ 
7 & 2.40  & 3.42  & 2.38  & 3.08 & 3.47 & 4.63  & 3.52 & 4.63  \\ 
8 & 2.55  & 3.69  & 2.51  & 3.35 & 3.55 &  5.17 & 3.49  & 5.17  \\ 
9 & 2.63 & 4.36  & 2.59 & 3.99 &  3.46 &  6.00 & 3.51  & 6.00  \\ 
10 & 3.87 & 6.04  & 3.83  & 5.58 &  5.23 & 7.71  & 5.34  & 7.71  \\ 
11 & 4.38 & 6.19  & 4.21  & 5.84 &  6.13 & 9.03 & 6.01 & 9.03  \\ 
12 & 5.46 & 7.40  & 5.25  & 7.40 & 8.27 &  11.78  & 8.03  & 11.78 \\ 
13 & 4.34 & 7.24  & 4.35  & 6.65 &  6.15 & 16.56  & 6.30 & 16.56  \\ 
14 & 4.68 & 8.20  & 5.26  & 8.24 &  8.63 &  15.28  & 9.11 & 15.28 \\ 
15 & 3.35 & 11.84  & 3.51  & 12.36 &  11.83  & 18.55 & 15.28 & 18.55  \\ 
\bottomrule
\end{tabular}
\end{table}

For various functional time-series models with the EVR criterion to select the number of components, we evaluate and compare their $\text{ECP}_h$, $\text{CPD}_h$ and $\overline{S}_{\alpha,h}$, where $h=1,\dots,15$. In Table~\ref{tab:2}, we present the averaged metrics $\overline{\text{ECP}}$, $\overline{\text{CPD}}$ and $\overline{S}_{\alpha}$, as well as the median M[ECP$_h$], M[CPD$_h$] and M[$\overline{S}_{\alpha,h}$]. At the $\alpha=0.2$ significance level, the conformal prediction interval approach coupled with the MLFTS generally provides the smallest mean and median CPD values and interval scores for both the CDF and CLR transformations.
\begin{small}
\begin{center}
\tabcolsep 0.16in
\renewcommand*{\arraystretch}{0.85}
\begin{longtable}{@{}lllrrr|rrr@{}}
\caption{\small At the nominal coverage probabilities of 80\%, we evaluate and compare the interval forecast accuracy between the conformal and standard deviation approaches, measured by $\text{ECP}$, $\text{CPD}$ and $S_{\alpha}$, for three functional time-series models with the EVR criterion for selecting the number of components. Based on the ECP and CPD, we highlight in bold the functional time-series method with the smallest values for each of the two approaches.}\label{tab:2} \\
\toprule
& & & \multicolumn{3}{c}{CDF}   & \multicolumn{3}{c}{CLR} \\
\cmidrule(lr){4-6}\cmidrule(lr){7-9}
Sex  & Metric & Approach & UFTS & MFTS & MLFTS & UFTS & MFTS & MLFTS \\ 
\midrule
\endfirsthead
\toprule
& & & \multicolumn{3}{c}{CDF}   & \multicolumn{3}{c}{CLR} \\
Sex & Metric & Approach & UFTS & MFTS & MLFTS & UFTS & MFTS & MLFTS \\ 
\midrule
\endhead
\hline \multicolumn{9}{r}{{Continued on next page}} \\
\endfoot
\endlastfoot
F 		& $\overline{\text{ECP}}$ 	& sd & 0.759 & 0.807 & 0.857 & 0.735 & 0.954 & 0.883  \\
				&					& conformal & 0.694 & 0.729 & 0.776 & 0.594 & 0.852 & 0.809 \\
\cmidrule{3-9}
	 	& M[ECP]				& sd & 0.757 & 0.840 & 0.838 & 0.707 & 0.985 & 0.905 \\
		&					&  conformal & 0.671 & 0.741 & 0.769 & 0.601 & 0.870 & 0.804 \\
\cmidrule{2-9}
	 	&  $\overline{\text{CPD}}$ & sd & 0.050 & \textBF{0.048} & 0.062 & 0.092 & 0.154 & \textBF{0.085} \\
		 		& 					& conformal & 0.106 & 0.071 & \textBF{0.037} & 0.206 & 0.055 & \textBF{0.047} \\
\cmidrule{3-9}
	 	& M[CPD]				& sd 	& 0.043 & 0.047 & \textBF{0.038} &	\textBF{0.093} & 0.185 & 0.105 \\
				&					& conformal 	& 0.129 & 0.059 & \textBF{0.032} & 0.199 & 0.070 & \textBF{0.047} \\
\cmidrule{2-9}
	 	&   $\overline{S}_{\alpha}$ & sd & 470.641 & 450.607 & \textBF{279.509} & 440.152 & 304.421 & \textBF{282.987} \\
				&					 & conformal & 447.700 & 430.650 & \textBF{285.221} & 422.189 & 308.855 & \textBF{258.568} \\
\cmidrule{3-9}
	 	& M[$S_{\alpha}$]		& sd 	& 480.484 & 407.819 & \textBF{249.753} & 475.842 & 280.564 & \textBF{262.005} \\
				&					& conformal 	& 460.567 & 400.257 & \textBF{270.113} & 438.396 & 319.200 & \textBF{234.998} \\
\midrule
	 M 	& $\overline{\text{ECP}}$ & sd & 0.845 & 0.702 & 0.802 & 0.950 & 0.662 & 0.833 \\
			& 					& conformal & 0.728 & 0.629 & 0.712 & 0.952 & 0.545 & 0.763 \\
\cmidrule{3-9}
	 	 		& M[ECP] 				& sd  & 0.855 & 0.685 & 0.790 & 0.947 & 0.665 & 0.839 \\
		 		& 					& conformal & 0.695 & 0.642 & 0.709 & 0.947 & 0.553 & 0.736 \\
\cmidrule{2-9}
	 	 		& $\overline{\text{CPD}}$ 	& sd & 0.053 & 0.098 & \textBF{0.032} & 0.150 & 0.138 & \textBF{0.080} \\
		 		& 					& conformal & \textBF{0.086} & 0.171 & 0.106 & 0.152 & 0.255 & \textBF{0.078} \\
\cmidrule{3-9}
	 	 		& M[CPD] 			& sd	& 0.055 & 0.115 & \textBF{0.027} & 0.147 & 0.135 & \textBF{0.069} \\
	 	 		& 					& conformal & 0.105 & 0.158 & \textBF{0.091} & 0.147 & 0.247 & \textBF{0.087} \\
\cmidrule{2-9}
	 	 		&   $\overline{S}_{\alpha}$ & sd & 324.628 & 297.966 & \textBF{286.152} & 520.602 & 406.046 & \textBF{291.875} \\
	 	 		& 					& conformal & 279.971 & 290.123 & \textBF{272.383} & 461.792 & 390.239 & \textBF{247.519} \\
\cmidrule{3-9}
	 	 		& M[$S_{\alpha}$] 		& sd	& 341.024 & 309.338 & \textBF{296.565} & 548.271 & 433.355 & \textBF{299.644} \\
	 	 		& 					&  conformal & 310.794 & 305.092 & \textBF{285.865} & 495.595 & 415.860 & \textBF{256.614} \\
\bottomrule
\end{longtable}
\end{center}
\end{small}

\vspace{-.3in}

For female data, under the CDF transformation, the MFTS (standard deviation approach) achieves the lowest mean CPD, while the MLFTS attains the lowest median CPD and the smallest mean and median interval scores. Under the CLR transformation, the UFTS minimizes median CPD but is slightly less effective than the MLFTS in mean CPD and interval scores.

For male data, under the CDF transformation, the MLFTS achieves the lowest mean and median CPD as well as the smallest interval scores. Under the CLR transformation, the MLFTS yields the smallest mean and median CPD. Taking into account the smallest mean and median interval scores, the MLFTS is the recommended choice.

At the $\alpha = 0.05$ significance level, Table~\ref{tab:3} highlights the conformal prediction interval with the MLFTS method as the best performer for female data under the CDF transformation. Under the CLR transformation, the MFTS method outperforms UFTS. For male data, MLFTS is recommended with the CLR transformation, while UFTS produces smaller CPD values and interval scores under the CDF transformation.
\begin{small}
\begin{center}
\tabcolsep 0.155in
\renewcommand{\arraystretch}{0.95}
\begin{longtable}{@{}lllrrr|rrr@{}}
\caption{\small At the nominal coverage probabilities of 95\%, we evaluate and compare the interval forecast accuracy between the conformal and standard deviation approaches, measured by $\text{ECP}_h$, $\text{CPD}_h$ and $S_{\alpha,h}$, for three functional time-series models with the EVR criterion for selecting the number of components.}\label{tab:3} \\
\toprule
& & & \multicolumn{3}{c}{CDF}   & \multicolumn{3}{c}{CLR} \\
\cmidrule(lr){4-6}\cmidrule(lr){7-9}
Sex & Metric & Approach & UFTS & MFTS & MLFTS & UFTS & MFTS & MLFTS \\ 
\midrule
\endfirsthead
\toprule
& & & \multicolumn{3}{c}{CDF}   & \multicolumn{3}{c}{CLR} \\
Sex & Metric & Approach & UFTS & MFTS & MLFTS & UFTS & MFTS & MLFTS \\ 
\midrule
\endhead
\hline \multicolumn{9}{r}{{Continued on next page}} \\
\endfoot
\endlastfoot
 F 		 	& $\overline{\text{ECP}}$ & sd & 0.856 & 0.872 & 0.948 & 0.854 & 0.988 & 0.954 \\
 				&					& conformal & 0.761 & 0.795 & 0.869 & 0.688 & 0.909 & 0.886 \\
\cmidrule{3-9}
				& M[ECP]				& sd & 0.868 & 0.885 & 0.958 & 0.857 & 0.999 & 0.962 \\
		 		& 					& conformal & 0.751 & 0.795 & 0.887 & 0.715 & 0.912 & 0.905 \\
\cmidrule{3-9}
	 	 		&  $\overline{\text{CPD}}$ & sd & 0.094 & 0.078 & \textBF{0.023} & 0.096 & 0.038 & \textBF{0.023} \\
		 		& 					& conformal & 0.189 & 0.155 & \textBF{0.081}& 0.262 & \textBF{0.041} & 0.064 \\
\cmidrule{3-9}
	 	 		& M[CPD]				& sd & 0.082 & 0.065 & \textBF{0.023} & 0.093 & 0.049 & \textBF{0.023} \\
		 		& 					& conformal & 0.199 & 0.155 & \textBF{0.063}& 0.235 & \textBF{0.038} & 0.045 \\
\cmidrule{3-9}
	 	 		&  $\overline{S}_{\alpha}$ & sd & 902.186 & 834.717 & \textBF{455.273} & 744.670 & 453.627 & \textBF{452.517} \\
		 		& 					& conformal & 1068.769 & 922.037 & \textBF{410.163} & 846.837 & 346.150 & \textBF{335.551}\\
\cmidrule{3-9}
	 	 		& M[$S_{\alpha}$]		& sd & 926.920 & 763.808 & \textBF{324.887} & 702.110 & 407.010 & \textBF{368.169} \\
		 		& 					& conformal & 1012.438 & 795.885 & \textBF{370.749} & 841.039 & \textBF{342.580} & 358.339 \\
\cmidrule{2-9}
 M 		&  $\overline{\text{ECP}}$ & sd & 0.946 & 0.886 & 0.921 & 0.977 & 0.873 & 0.939 \\
 		& 					& conformal & 0.798 & 0.718 & 0.784 & 0.975 & 0.636 & 0.853 \\
\cmidrule{3-9}
	 	 		& M[ECP]	& sd & 0.953 & 0.873 & 0.908 & 0.985 & 0.867 & 0.929 \\
		 		& 		& conformal & 0.743 & 0.730 & 0.755 & 0.980 & 0.637 & 0.851 \\
\cmidrule{3-9}
	 	 		&  $\overline{\text{CPD}}$ & sd & \textBF{0.018} & 0.068 & 0.040  & \textBF{0.031} & 0.078 & 0.038 \\
		 		& 					& conformal & \textBF{0.152} & 0.232 & 0.167 & \textBF{0.026} & 0.314 & 0.097\\ 
\cmidrule{3-9}
	 	 		& M[CPD]	 & sd & \textBF{0.014} & 0.077 & 0.042 & \textBF{0.035} & 0.083 & 0.036 \\
		 		& 		& conformal & 0.207 & 0.220 & \textBF{0.195} & \textBF{0.030} & 0.313 & 0.099 \\
\cmidrule{3-9}
	 	 		&  $\overline{S}_{\alpha}$ & sd & 493.131 & 475.002 & \textBF{456.609} & 660.490 & 611.946 & \textBF{460.363} \\
		 		& 					& conformal & \textBF{422.177} & 579.461 & 501.265 & 522.754 & 839.853 & \textBF{368.778} \\
\cmidrule{3-9}
	 	 		& M[$S_{\alpha}$]	& sd & 448.888 & \textBF{425.326} & 445.072 & 750.222 & 552.793 & \textBF{447.374} \\
		 		& 				& conformal & \textBF{457.082} & 572.234 & 483.798 & 579.485 & 848.386 & \textBF{374.045} \\
\bottomrule
\end{longtable}
\end{center}
\end{small}
  
\vspace{-.4in}
  
For female data, MLFTS achieves the lowest mean and median CPD, along with the smallest interval scores, under both the CDF and CLR transformations. For male data, the UFTS yields the lowest mean and median CPD, although the MLFTS and MFTS provide better interval scores. Under the CLR transformation, the UFTS attains the smallest mean and median CPD, but considering interval scores, the MLFTS remains the preferred choice. For comparison, we also consider $K=6$ number of components and report their results in the Appendix~A.

\section{Conclusion}\label{sec:5}

We propose a general strategy for constructing prediction intervals for the age distribution of death. This approach leverages a validation set to determine an optimal tuning parameter that aligns empirical and nominal coverage probabilities. Using this optimized parameter, we construct prediction intervals for the testing set.

To illustrate the effectiveness of this strategy, we analyze Japanese age- and sex-specific life-table death counts, comparing three functional time-series forecasting models: univariate, multivariate, and multilevel functional time-series models. Our findings suggest that the multilevel functional time-series method generally performs best. Additionally, when selecting the number of components, we find little difference between the EVR criterion and the setting $K=6$. Given that overfitting does not adversely affect the accuracy of the forecast, we recommend the latter.

Using the age distribution of deaths, we present our methodology for constructing distribution-free and model-agnostic prediction intervals. Other measures of mortality, such as age-specific mortality rates or hazard rates, could also be considered in the modeling. In Appendixes~B and~C, we demonstrate our proposed sd approach for constructing prediction intervals and evaluating its empirical coverage probability using the Australian and Canadian age- and sex-specific $\log$ mortality rates, respectively. Using the proposed sd approach, it achieves superior finite-sample coverage probability in comparison to the classical functional time-series model of \cite{HU07}. The method of \cite{HU07} computes the total variance and constructs the prediction interval parametrically based on the assumption of a Gaussian distribution.

There are at least five ways in which the methodology can be extended.
\begin{inparaenum}
\item[1)] The functional standard deviation was computed coordinate-wise. Several functional depths exist, which can be implemented to compute other variants of standard deviations.
\item[2)] Instead of symmetric prediction intervals, one can consider asymmetric ones. In that case, two tuning parameters are needed to adjust the lower and upper bounds. 
\item[3)] The data set was divided equally into training, validation, and testing samples. Other proportions may be possible and lead to a more accurate selection of the tuning parameter $\xi_{\alpha}$ and more accurate interval forecasts.
\item[4)] For demonstration, we implemented a suite of functional time-series models. Other time-series extrapolation models may also be considered.
\item[5)] We use the life-table data directly in our modeling, but we could extend the analysis by incorporating their estimation error into the model, reflecting the underlying observational data used to contruct the life table.
\end{inparaenum}

\section*{Acknowledgment}

The authors are grateful for several motivating suggestions of one anonymous reviewer. The first author thanks funding from the Australian Research Council Discovery Project DP230102250 and Future Fellowship FT240100338.

\section*{Appendix A: Interval forecast results when $K=6$}

In Tables~\ref{tab:S_1} and~\ref{tab:S_2}, we present the interval forecast accuracy between the conformal and standard deviation approaches for three functional time-series models with the first six retained components at the nominal coverage probabilities of 80\% and 95\%, respectively.

From Table~\ref{tab:S_1}, the MFTS and MLFTS outperform the UFTS with smaller mean and median CPD and interval scores for both approaches. However, for male data, the UFTS achieves a lower CPD using the standard deviation approach with the CLR transformation.
\begin{small}
\begin{center}
\tabcolsep 0.16in
\renewcommand{\arraystretch}{0.92}
\begin{longtable}{@{}lllrrr|rrr@{}}
\caption{\small At the nominal coverage probabilities of 80\%, we evaluate and compare the interval forecast accuracy between the conformal and standard deviation approaches, measured by $\text{ECP}$, $\text{CPD}$ and $S_{\alpha}$, for three functional time-series models with $K=6$. Based on the ECP and CPD, we highlight in bold the functional time-series method with the smallest values for each of the two approaches.}\label{tab:S_1} \\
\toprule
&  & & \multicolumn{3}{c}{CDF}   & \multicolumn{3}{c}{CLR} \\
 Sex &  Metric & Approach & UFTS & MFTS & MLFTS & UFTS & MFTS & MLFTS \\ 
\midrule
\endfirsthead
\toprule
 & & & \multicolumn{3}{c}{CDF}   & \multicolumn{3}{c}{CLR} \\
 Sex & Metric & Approach & UFTS & MFTS & MLFTS & UFTS & MFTS & MLFTS \\ 
\midrule
\endhead
\hline \multicolumn{9}{r}{{Continued on next page}} \\
\endfoot
\endlastfoot
F 		 	& $\overline{\text{ECP}}$ 	& sd & 0.750 & 0.801 & 0.833 & 0.785 & 0.835 & 0.878 \\ 
			&					& conformal & 0.681 & 0.726 & 0.766 & 0.654 & 0.815 & 0.753 \\ 
\cmidrule{3-9}
	 		& M[ECP]				& sd & 0.746 & 0.809 & 0.832 & 0.771 & 0.821 & 0.873 \\ 
				&					&  conformal & 0.662 & 0.726 & 0.764 & 0.677 & 0.825 & 0.755 \\ 
\cmidrule{2-9}
	 	 		&  $\overline{\text{CPD}}$ & sd & 0.061 & \textBF{0.036} & 0.037 & \textBF{0.050} & 0.071 & 0.083 \\ 
		 		& 					& conformal  & 0.119 & 0.074 & \textBF{0.035} & 0.146 & 0.065 & \textBF{0.061} \\ 
\cmidrule{3-9}
	 	 		& M[CPD]				& sd 	& 0.054 & 0.035 & \textBF{0.034} & \textBF{0.042} & 0.066 & 0.073 \\
				&					& conformal 	& 0.138 & 0.074 & \textBF{0.036} & 0.123 & 0.072 & \textBF{0.045} \\ 
\cmidrule{2-9}
	 	 		&   $\overline{S}_{\alpha}$ & sd  & 425.527 & 336.788 & \textBF{309.193} & 376.570 & 309.369 & \textBF{274.407} \\  
				&					 & conformal  & 404.812 & 327.194 & \textBF{311.363} & 351.138 & \textBF{256.756} & 259.693 \\ 
\cmidrule{3-9}
	 	 		& M[$S_{\alpha}$]		& sd 	& 422.625 & 318.620 & \textBF{264.745} & 362.494 & 281.715 & \textBF{232.967} \\ 			
				&					& conformal     & 408.239 & 324.021 & \textBF{293.286} & 367.602 & \textBF{265.018} & 270.223 \\ 	 
\cmidrule{2-9}
	 M 	&	 $\overline{\text{ECP}}$ & sd & 0.815 & 0.737 & 0.825 & 0.854 & 0.747 & 0.887 \\
			& 					& conformal & 0.701 & 0.652 & 0.740 & 0.768 & 0.652 & 0.810 \\ 
\cmidrule{3-9}
	 	 		& M[ECP] 				& sd  & 0.820 & 0.735 & 0.820 & 0.841 & 0.733 & 0.875 \\
		 		& 					& conformal & 0.669 & 0.655 & 0.728 & 0.783 & 0.646 & 0.819 \\ 
\cmidrule{2-9}
	 	 		& $\overline{\text{CPD}}$ 	& sd & 0.035 & 0.063 & \textBF{0.026} & \textBF{0.057} & 0.058 & 0.087 \\
		 		& 					& conformal & 0.101 & 0.148 & \textBF{0.068} & 0.066 & 0.148 & \textBF{0.025} \\ 
\cmidrule{3-9}
	 	 		& M[CPD] 			& sd	& 0.034 & 0.065 & \textBF{0.020} & \textBF{0.041} & 0.067 & 0.075 \\
	 	 		& 					& conformal  & 0.131 & 0.145 & \textBF{0.072} & 0.049 & 0.154 & \textBF{0.021} \\  
\cmidrule{2-9}
	 	 		&   $\overline{S}_{\alpha}$ & sd  & 332.674 & 318.413 & \textBF{289.418} & 336.100 & 444.925 & \textBF{275.450} \\
	 	 		& 					& conformal & 285.552 & 302.708 & \textBF{263.287} & 279.133 & 406.393 & \textBF{234.783} \\ 
\cmidrule{3-9}
	 	 		& M[$S_{\alpha}$] 		& sd	& 354.513 & 347.581 & \textBF{331.476} & 338.271 & 457.361 & \textBF{289.172} \\	
	 	 		& 					&  conformal  & 327.190 & 308.440 & \textBF{298.650} & 274.904 & 448.065 & \textBF{257.413} \\
\bottomrule
\end{longtable}
\end{center}
\end{small}

\vspace{-.3in}

From Table~\ref{tab:S_2}, the MFTS and MLFTS outperform the UFTS with smaller mean and median CPD and interval scores for both approaches. However, for male data, the UFTS achieves a lower CPD and an interval score using the standard deviation approach with the CDF transformation.
\begin{small}
\begin{center}
\tabcolsep 0.16in
\renewcommand{\arraystretch}{0.93}
\begin{longtable}{@{}lllrrr|rrr@{}}
\caption{\small At the nominal coverage probabilities of 95\%, we evaluate and compare the interval forecast accuracy between the conformal and standard deviation approaches, measured by $\text{ECP}_h$, $\text{CPD}_h$ and $S_{\alpha,h}$, for three functional time-series models with $K=6$.}\label{tab:S_2} \\
\toprule
&  & & \multicolumn{3}{c}{CDF}   & \multicolumn{3}{c}{CLR} \\
 Sex & Metric & Approach & UFTS & MFTS & MLFTS & UFTS & MFTS & MLFTS \\ 
\midrule
\endfirsthead
\toprule
& & & \multicolumn{3}{c}{CDF}   & \multicolumn{3}{c}{CLR} \\
 Sex  & Metric & Approach & UFTS & MFTS & MLFTS & UFTS & MFTS & MLFTS \\ 
\midrule
\endhead
\hline \multicolumn{9}{r}{{Continued on next page}} \\
\endfoot
\endlastfoot
 F 		& $\overline{\text{ECP}}$ & sd & 0.859 & 0.891 & 0.931 &  0.900 & 0.924 & 0.963 \\
 				&					& conformal & 0.755 & 0.801 & 0.839 & 0.745 & 0.887 & 0.833 \\ 
\cmidrule{3-9}
		 		& M[ECP]				& sd & 0.865 & 0.884 & 0.929 & 0.892 & 0.923 & 0.971 \\
		 		& 					& conformal  & 0.742 & 0.806 & 0.824 &  0.776 & 0.868 & 0.841 \\ 
\cmidrule{3-9}
	 	 		&  $\overline{\text{CPD}}$ & sd  & 0.091 & 0.059 & \textBF{0.023} & 0.050 & 0.048 & \textBF{0.022} \\
		 		& 					& conformal & 0.195 & 0.149 & \textBF{0.111} & 0.205 & \textBF{0.064} & 0.117 \\ 
\cmidrule{3-9}
	 	 		& M[CPD]				& sd & 0.085 & 0.066 & \textBF{0.023} & 0.058 & 0.042 & \textBF{0.021} \\
		 		& 					& conformal &0.208 & 0.144 & \textBF{0.126} & 0.174 & \textBF{0.082} & 0.109 \\ 
\cmidrule{3-9}
	 	 		&  $\overline{S}_{\alpha}$ & sd & 783.946 & 582.336 & \textBF{491.354} & 613.577 & 557.077 & \textBF{434.461} \\
		 		& 					& conformal  & 943.683 & 652.737 & \textBF{574.971} & 606.063 & 379.501 & \textBF{341.970} \\ 
\cmidrule{3-9}
	 	 		& M[$S_{\alpha}$]		& sd  & 732.045 & 510.697 & \textBF{371.856} & 532.389 & 445.167 & \textBF{326.626} \\	
		 		& 					& conformal & 852.400 & 573.037 & \textBF{437.359} & 521.617 & \textBF{343.097} & 358.741 \\ 
\cmidrule{2-9}
 M 		 	&  $\overline{\text{ECP}}$ & sd & 0.946 & 0.882 & 0.937 & 0.942 & 0.867 & 0.961 \\
	 	 		& 					& conformal & 0.771 & 0.742 & 0.814 & 0.848 & 0.749 & 0.877 \\ 
\cmidrule{3-9}
	 	 		& M[ECP]	& sd & 0.953 & 0.872 & 0.938 & 0.945 & 0.864 & 0.956 \\
		 		& 		& conformal & 0.717 & 0.744 & 0.786 & 0.851 & 0.747 & 0.886 \\  
\cmidrule{3-9}
	 	 		&  $\overline{\text{CPD}}$ & sd & \textBF{0.018} & 0.068 & 0.022 & 0.030 & 0.083 & \textBF{0.015} \\
		 		& 					& conformal  & 0.179 & 0.208 & \textBF{0.136} & 0.102 & 0.201 & \textBF{0.073} \\ 
\cmidrule{3-9}
	 	 		& M[CPD]	 & sd & \textBF{0.014} & 0.078 & 0.019 & 0.027 & 0.086 & \textBF{0.008} \\
		 		& 		& conformal & 0.233 & 0.206 & \textBF{0.164} & 0.099 & 0.203 & \textBF{0.064} \\ 
\cmidrule{3-9}
	 	 		&  $\overline{S}_{\alpha}$ & sd  & \textBF{493.131} & 602.081 & 493.277 & 662.022 & 762.599 & \textBF{462.813} \\
		 		& 					& conformal & 454.399 & 632.817 & \textBF{418.730} & 442.703 & 874.558 & \textBF{327.106} \\  
\cmidrule{3-9}
	 	 		& M[$S_{\alpha}$]	& sd & \textBF{448.888} & 562.958 & 477.297 & 575.785 & 836.510 & \textBF{408.421} \\
		 		& 				& conformal & 518.473 & 639.114 & \textBF{434.471} & 354.886 & 886.527 & \textBF{333.583} \\ 
\bottomrule
\end{longtable}
\end{center}
\end{small}
  
\newpage
\subsection*{Appendix B: Australian age-specific mortality rates}
  
\begin{figure}[!htb]
\centering
{\includegraphics[width=8.4cm]{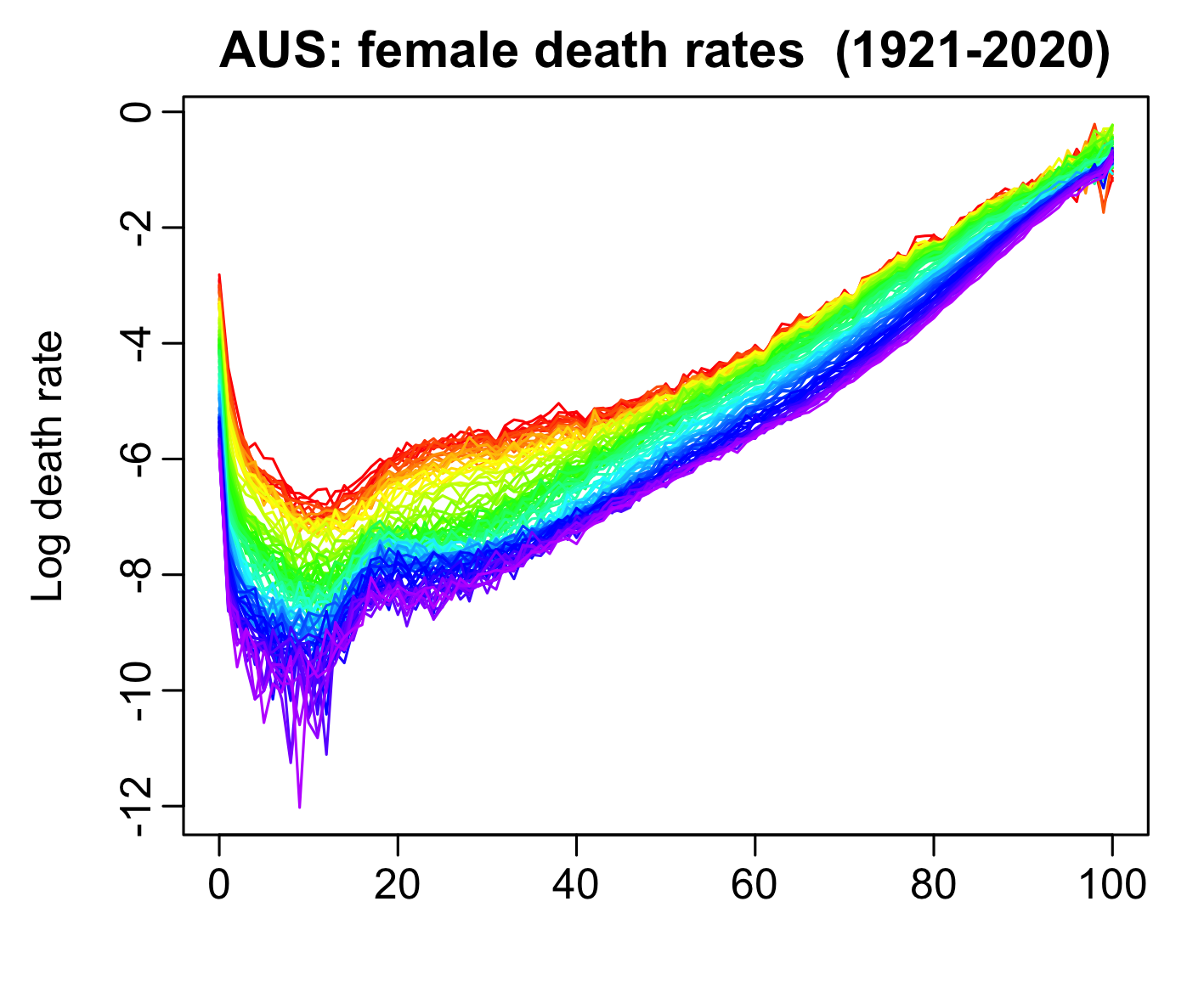}}
\quad
{\includegraphics[width=8.4cm]{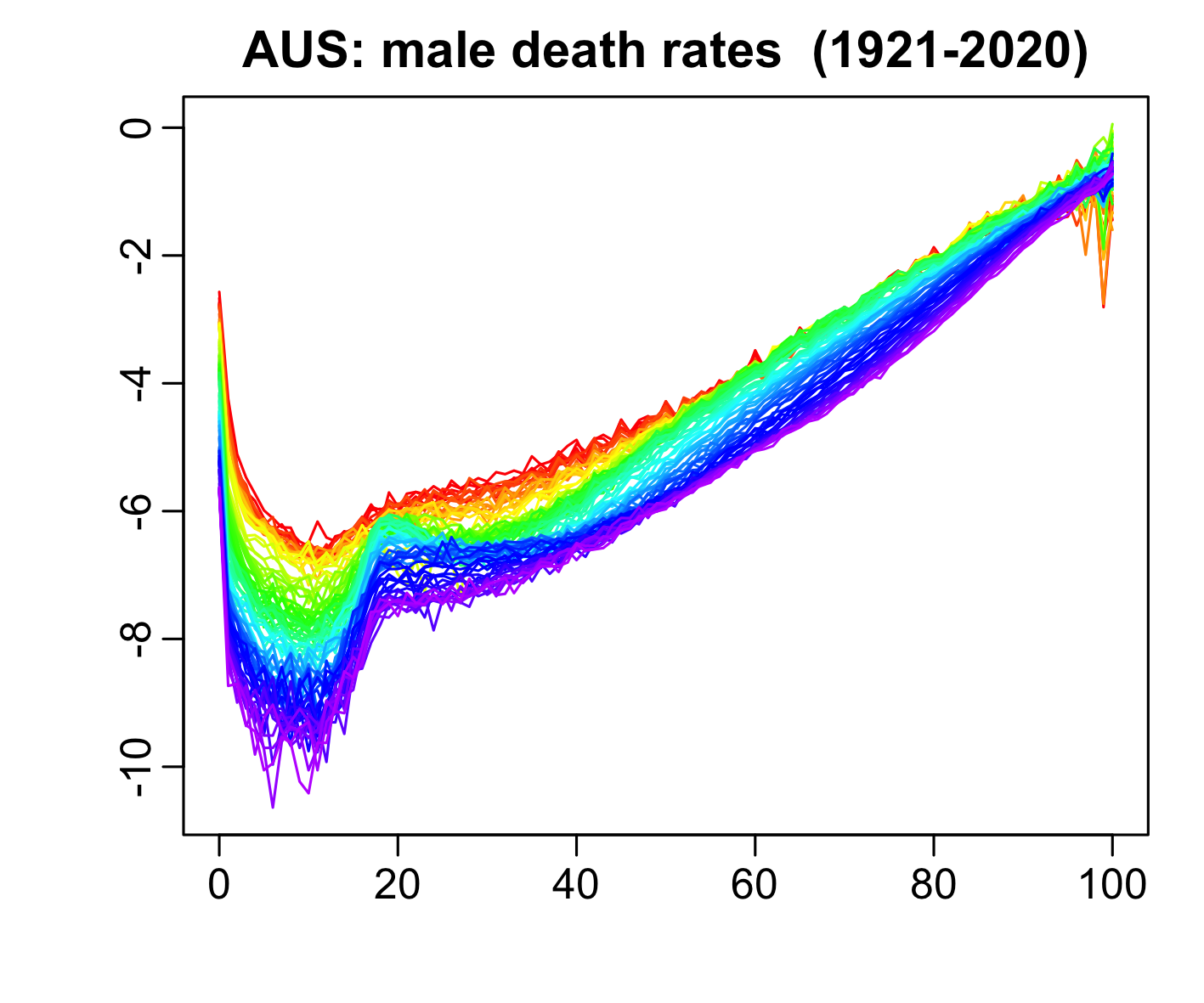}}
\\
{\includegraphics[width=8.4cm]{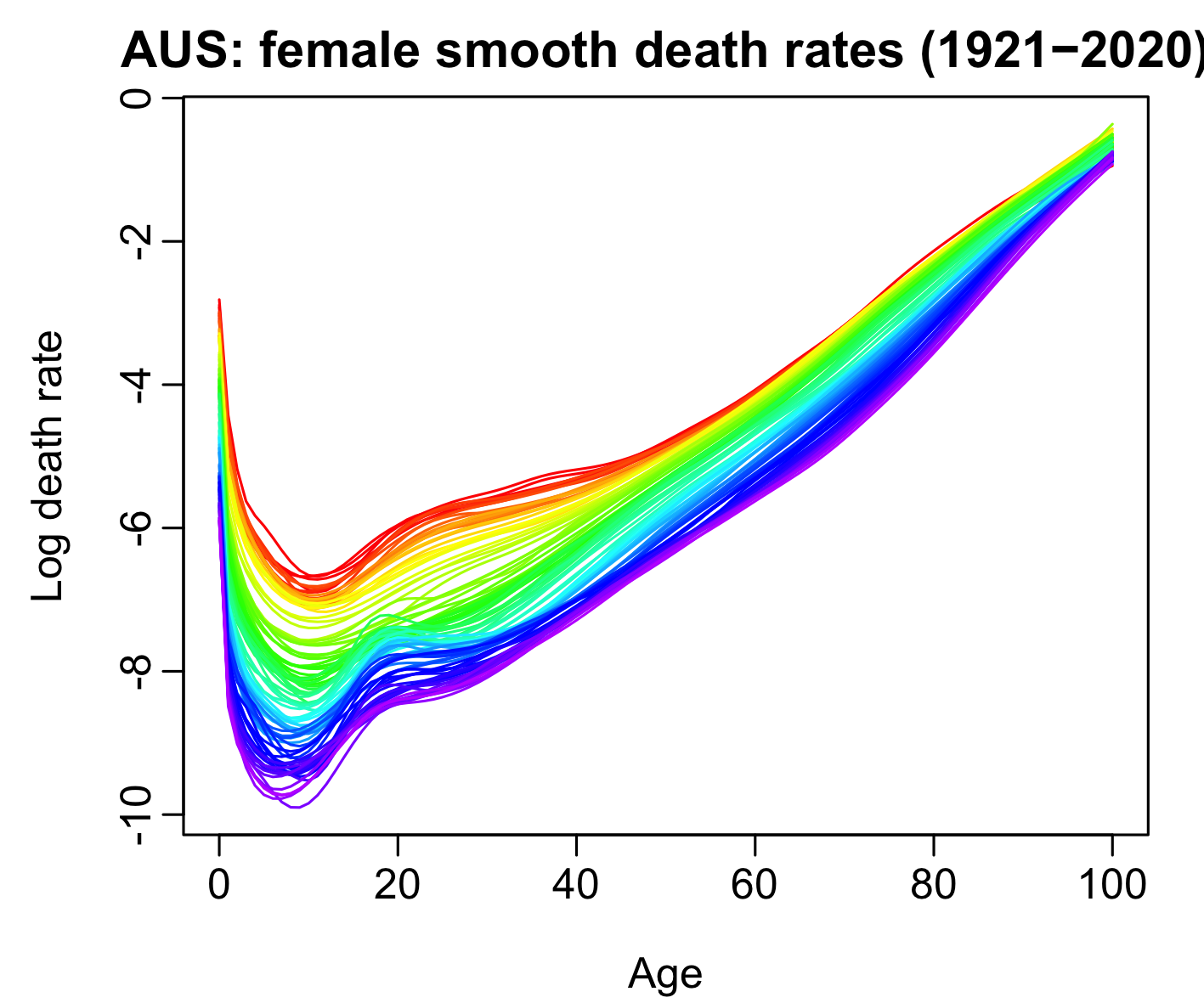}}
\quad
{\includegraphics[width=8.4cm]{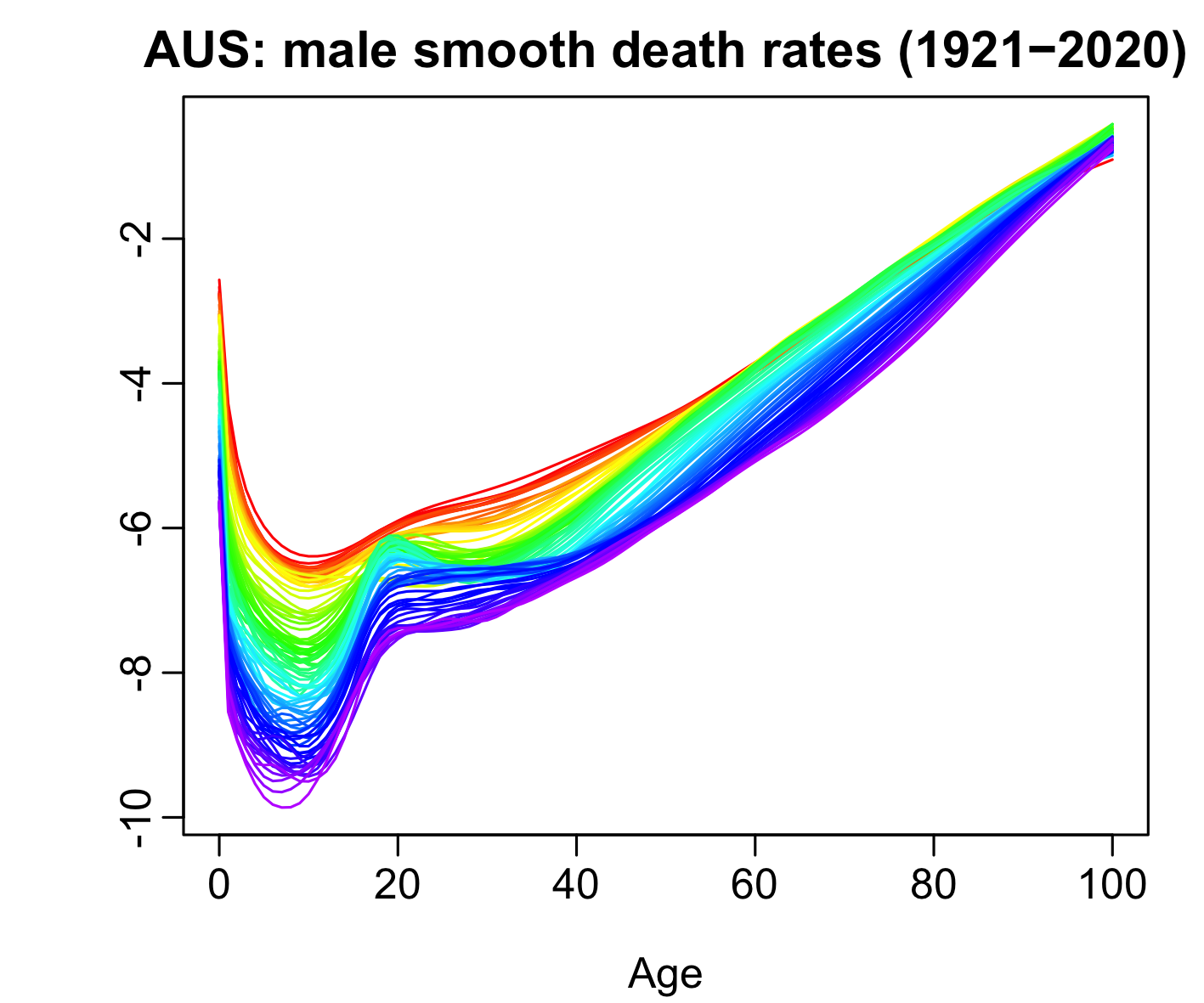}}
\caption{\small Rainbow plots of the original and smoothed age-specific mortality rates for the Australian female and male data from 1921 to 2020. Smoothing was performed via penalized spline with monotonic constraint described in \cite{HU07}.}\label{fig:A_1}
\end{figure}

We analyze Australian age- and sex-specific mortality rates spanning from 1921 to 2020, obtained from \cite{HMD24}. These rates represent the ratio of death counts to population exposure in each respective year and age group (based on one-year intervals). Our study covers age groups from 0 to 99 in single years, with the final group covering ages 100 and above. Age-specific mortality rates are often modeled and forecasted using natural logarithmic transformations. In Figure~\ref{fig:A_1}, we present rainbow plots for $\log$ mortality rates.
  
In Table~\ref{tab:A_1}, we compute the empirical coverage probability specific to each horizon and its coverage probability difference for the sd approach at the nominal coverage probability 80\%. For comparison, we implement the parametric approach of \cite{HU07} implemented in the ftsa package in \Rlogo. Using data from 1921 to 1976, we computed the forecasts for the validation period from 1977 to 1998. For $h=1, 2,\dots,21$, we determine the optimal tuning parameters, with which we evaluate the empirical coverage probability based on the test period from 1999 to 2020. The proposed sd approach achieves superior finite-sample coverage in comparison to the parametric approach based on the total variance under the Gaussian distribution assumption.
\begin{table}[!htb]
\tabcolsep 0.04in
\renewcommand{\arraystretch}{0.85}
\centering
\caption{\small At the nominal coverage probability of 80\%, we compute the empirical coverage probability and its coverage probability difference between the sd approach and parametric approach.}\label{tab:A_1}
\begin{small}
\begin{tabular}{@{}lrrrrrrrrrrrrrrrr@{}}
\toprule
& \multicolumn{8}{c}{sd approach}   & \multicolumn{8}{c}{parametric approach} \\
  \cmidrule(lr){2-9}  \cmidrule(lr){10-17}
& \multicolumn{4}{c}{Female} & \multicolumn{4}{c}{Male}  & \multicolumn{4}{c}{Female} & \multicolumn{4}{c}{Male} \\
\cmidrule(lr){2-5}\cmidrule(lr){6-9}\cmidrule(lr){10-13}\cmidrule(lr){14-17}	
& \multicolumn{2}{c}{Smooth} 	& \multicolumn{2}{c}{Raw}  & \multicolumn{2}{c}{Smooth} & \multicolumn{2}{c}{Raw} & \multicolumn{2}{c}{Smooth} 	& \multicolumn{2}{c}{Raw}  & \multicolumn{2}{c}{Smooth} 	& \multicolumn{2}{c}{Raw}  \\ 
\midrule
$h$ & ECP & CPD & ECP & CPD & ECP & CPD & ECP & CPD & ECP & CPD & ECP & CPD & ECP & CPD & ECP & CPD \\ 
\midrule
1 & 0.807 & 0.007 & 0.802 & 0.002 & 0.815 & 0.015 & 0.849 & 0.049 & 0.545 & 0.255 & 0.746 & 0.054 & 0.573 & 0.227 & 0.738 & 0.062 \\ 
2 & 0.819 & 0.019 & 0.836 & 0.036 & 0.846 & 0.046 & 0.849 & 0.049 & 0.544 & 0.256 & 0.749 & 0.051 & 0.529 & 0.271 & 0.686 & 0.114 \\  
3 & 0.815 & 0.015 & 0.838 & 0.038 & 0.875 & 0.075 & 0.875 & 0.075 & 0.552 & 0.248 & 0.739 & 0.061 & 0.498 & 0.302 & 0.645 & 0.155 \\ 
4 & 0.829 & 0.029 & 0.845 & 0.045 & 0.903 & 0.103 & 0.900 & 0.100 & 0.553 & 0.247 & 0.730 & 0.070 & 0.476 & 0.324 & 0.594 & 0.206 \\  
5 & 0.824 & 0.024 & 0.842 & 0.042 & 0.923 & 0.123 & 0.926 & 0.126 & 0.558 & 0.242 & 0.726 & 0.074 & 0.450 & 0.350 & 0.563 & 0.237 \\  
6 & 0.832 & 0.032 & 0.856 & 0.056 & 0.949 & 0.149 & 0.945 & 0.145 & 0.562 & 0.238 & 0.715 & 0.085 & 0.425 & 0.375 & 0.533 & 0.267 \\  
7 & 0.822 & 0.022 & 0.853 & 0.053 & 0.954 & 0.154 & 0.957 & 0.157 & 0.544 & 0.256 & 0.697 & 0.103 & 0.405 & 0.395 & 0.502 & 0.298 \\  
8 & 0.825 & 0.025 & 0.861 & 0.061 & 0.980 & 0.180 & 0.983 & 0.183 & 0.547 & 0.253 & 0.700 & 0.100 & 0.389 & 0.411 & 0.496 & 0.304 \\  
9 & 0.822 & 0.022 & 0.873 & 0.073 & 0.993 & 0.193 & 0.988 & 0.188 & 0.557 & 0.243 & 0.690 & 0.110 & 0.380 & 0.420 & 0.497 & 0.303 \\  
10 & 0.819 & 0.019 & 0.881 & 0.081 & 0.989 & 0.189 & 0.984 & 0.184 & 0.560 & 0.240 & 0.691 & 0.109 & 0.369 & 0.431 & 0.486 & 0.314 \\  
11 & 0.830 & 0.030 & 0.870 & 0.070 & 0.979 & 0.179 & 0.979 & 0.179 & 0.547 & 0.253 & 0.676 & 0.124 & 0.371 & 0.429 & 0.481 & 0.319 \\  
12 & 0.833 & 0.033 & 0.887 & 0.087 & 0.977 & 0.177 & 0.984 & 0.184 & 0.557 & 0.243 & 0.671 & 0.129 & 0.364 & 0.436 & 0.466 & 0.334 \\  
13 & 0.849 & 0.049 & 0.885 & 0.085 & 0.983 & 0.183 & 0.990 & 0.190 & 0.566 & 0.234 & 0.673 & 0.127 & 0.366 & 0.434 & 0.471 & 0.329 \\  
14 & 0.827 & 0.027 & 0.914 & 0.114 & 0.983 & 0.183 & 0.985 & 0.185 & 0.554 & 0.246 & 0.673 & 0.127 & 0.372 & 0.428 & 0.483 & 0.317 \\  
15 & 0.830 & 0.030 & 0.901 & 0.101 & 0.978 & 0.178 & 0.978 & 0.178 & 0.561 & 0.239 & 0.663 & 0.137 & 0.395 & 0.405 & 0.459 & 0.341 \\  
16 & 0.789 & 0.011 & 0.881 & 0.081 & 0.969 & 0.169 & 0.963 & 0.163 & 0.549 & 0.251 & 0.652 & 0.148 & 0.402 & 0.398 & 0.463 & 0.337 \\  
17 & 0.807 & 0.007 & 0.875 & 0.075 & 0.969 & 0.169 & 0.959 & 0.159 & 0.540 & 0.260 & 0.645 & 0.155 & 0.408 & 0.392 & 0.464 & 0.336 \\  
18 & 0.794 & 0.006 & 0.861 & 0.061 & 0.970 & 0.170 & 0.956 & 0.156 & 0.507 & 0.293 & 0.650 & 0.150 & 0.420 & 0.380 & 0.461 & 0.339 \\  
19 & 0.829 & 0.029 & 0.869 & 0.069 & 0.978 & 0.178 & 0.975 & 0.175 & 0.505 & 0.295 & 0.641 & 0.159 & 0.426 & 0.374 & 0.475 & 0.325 \\  
20 & 0.825 & 0.025 & 0.931 & 0.131 & 0.927 & 0.127 & 0.921 & 0.121 & 0.492 & 0.308 & 0.647 & 0.153 & 0.419 & 0.381 & 0.488 & 0.312 \\  
21 & 0.856 & 0.056 & 0.931 & 0.131 & 0.876 & 0.076 & 0.921 & 0.121 & 0.550 & 0.250 & 0.673 & 0.127 & 0.426 & 0.374 & 0.485 & 0.315 \\ 
\midrule
Mean & 0.823 & 0.025 & 0.871 & 0.071 & 0.944 & 0.144 & 0.946 & 0.146 & 0.545 & 0.255 & 0.688 & 0.112 & 0.422 & 0.378 & 0.521 & 0.279 \\
\bottomrule
\end{tabular}
\end{small}  
\end{table}

In Table~\ref{tab:A_2}, we report the empirical coverage probability specific to each horizon and its coverage probability difference for the sd approach at the nominal level 95\%. Compared with the parametric approach of \cite{HU07}, the sd approach achieves superior finite-sample coverage probability and often produces empirical coverage probability at and above the nominal level.
\begin{table}[!htb]
\tabcolsep 0.04in
\renewcommand{\arraystretch}{0.93}
\centering
\caption{\small At the nominal coverage probability of 95\%, we compute the empirical coverage probability and its coverage probability difference between the sd approach and parametric approach.}\label{tab:A_2}
\begin{small}
\begin{tabular}{@{}lrrrrrrrrrrrrrrrr@{}}
\toprule
& \multicolumn{8}{c}{sd approach}   & \multicolumn{8}{c}{parametric approach} \\
  \cmidrule(lr){2-9}  \cmidrule(lr){10-17}
& \multicolumn{4}{c}{Female} & \multicolumn{4}{c}{Male}  & \multicolumn{4}{c}{Female} & \multicolumn{4}{c}{Male} \\
\cmidrule(lr){2-5}\cmidrule(lr){6-9}\cmidrule(lr){10-13}\cmidrule(lr){14-17}	
& \multicolumn{2}{c}{Smooth} 	& \multicolumn{2}{c}{Raw}  & \multicolumn{2}{c}{Smooth} 	& \multicolumn{2}{c}{Raw} & \multicolumn{2}{c}{Smooth} 	& \multicolumn{2}{c}{Raw}  & \multicolumn{2}{c}{Smooth} 	& \multicolumn{2}{c}{Raw}  \\ 
\midrule
$h$ & ECP & CPD & ECP & CPD & ECP & CPD & ECP & CPD & ECP & CPD & ECP & CPD & ECP & CPD & ECP & CPD \\ 
\midrule
1 & 0.933 & 0.017 & 0.938 & 0.012 & 0.949 & 0.001 & 0.964 & 0.014 & 0.715 & 0.235 & 0.907 & 0.043 & 0.752 & 0.198 & 0.899 & 0.051 \\ 
2 & 0.938 & 0.012 & 0.943 & 0.007 & 0.972 & 0.022 & 0.980 & 0.030 & 0.722 & 0.228 & 0.906 & 0.044 & 0.736 & 0.214 & 0.883 & 0.067 \\  
3 & 0.943 & 0.007 & 0.952 & 0.002 & 0.973 & 0.023 & 0.982 & 0.032 & 0.722 & 0.228 & 0.905 & 0.045 & 0.707 & 0.243 & 0.848 & 0.102 \\ 
4 & 0.947 & 0.003 & 0.957 & 0.007 & 0.981 & 0.031 & 0.984 & 0.034 & 0.741 & 0.209 & 0.899 & 0.051 & 0.688 & 0.262 & 0.826 & 0.124 \\ 
5 & 0.952 & 0.002 & 0.964 & 0.014 & 0.989 & 0.039 & 0.988 & 0.038 & 0.737 & 0.213 & 0.893 & 0.057 & 0.644 & 0.306 & 0.806 & 0.144 \\  
6 & 0.944 & 0.006 & 0.956 & 0.006 & 0.991 & 0.041 & 0.997 & 0.047 & 0.732 & 0.218 & 0.888 & 0.062 & 0.620 & 0.330 & 0.775 & 0.175 \\  
7 & 0.940 & 0.010 & 0.958 & 0.008 & 0.993 & 0.043 & 0.991 & 0.041 & 0.727 & 0.223 & 0.876 & 0.074 & 0.592 & 0.358 & 0.751 & 0.199 \\  
8 & 0.947 & 0.003 & 0.964 & 0.014 & 0.999 & 0.049 & 0.997 & 0.047 & 0.723 & 0.227 & 0.884 & 0.066 & 0.585 & 0.365 & 0.739 & 0.211 \\  
9 & 0.937 & 0.013 & 0.964 & 0.014 & 1.000 & 0.050 & 1.000 & 0.050 & 0.723 & 0.227 & 0.883 & 0.067 & 0.574 & 0.376 & 0.732 & 0.218 \\  
10 & 0.941 & 0.009 & 0.966 & 0.016 & 1.000 & 0.050 & 1.000 & 0.050 & 0.717 & 0.233 & 0.858 & 0.092 & 0.561 & 0.389 & 0.720 & 0.230 \\  
11 & 0.947 & 0.003 & 0.972 & 0.022 & 0.998 & 0.048 & 0.993 & 0.043 & 0.715 & 0.235 & 0.861 & 0.089 & 0.568 & 0.382 & 0.705 & 0.245 \\  
12 & 0.945 & 0.005 & 0.977 & 0.027 & 0.995 & 0.045 & 1.000 & 0.050 & 0.698 & 0.252 & 0.849 & 0.101 & 0.557 & 0.393 & 0.697 & 0.253 \\ 
13 & 0.944 & 0.006 & 0.977 & 0.027 & 0.995 & 0.045 & 0.999 & 0.049 & 0.691 & 0.259 & 0.846 & 0.104 & 0.574 & 0.376 & 0.693 & 0.257 \\  
14 & 0.944 & 0.006 & 0.981 & 0.031 & 0.994 & 0.044 & 0.994 & 0.044 & 0.691 & 0.259 & 0.838 & 0.112 & 0.582 & 0.368 & 0.691 & 0.259 \\  
15 & 0.965 & 0.015 & 0.984 & 0.034 & 0.996 & 0.046 & 0.990 & 0.040 & 0.689 & 0.261 & 0.832 & 0.118 & 0.571 & 0.379 & 0.676 & 0.274 \\  
16 & 0.949 & 0.001 & 0.987 & 0.037 & 0.992 & 0.042 & 0.982 & 0.032 & 0.679 & 0.271 & 0.826 & 0.124 & 0.576 & 0.374 & 0.668 & 0.282 \\  
17 & 0.957 & 0.007 & 0.975 & 0.025 & 0.995 & 0.045 & 0.990 & 0.040 & 0.685 & 0.265 & 0.820 & 0.130 & 0.576 & 0.374 & 0.660 & 0.290 \\  
18 & 0.937 & 0.013 & 0.978 & 0.028 & 0.996 & 0.046 & 0.980 & 0.030 & 0.685 & 0.265 & 0.822 & 0.128 & 0.560 & 0.390 & 0.659 & 0.291 \\  
19 & 0.955 & 0.005 & 0.980 & 0.030 & 1.000 & 0.050 & 1.000 & 0.050 & 0.671 & 0.279 & 0.832 & 0.118 & 0.554 & 0.396 & 0.661 & 0.289 \\  
20 & 0.944 & 0.006 & 0.990 & 0.040 & 0.990 & 0.040 & 0.990 & 0.040 & 0.677 & 0.273 & 0.805 & 0.145 & 0.548 & 0.402 & 0.644 & 0.306 \\  
21 & 0.975 & 0.025 & 0.970 & 0.020 & 0.965 & 0.015 & 0.990 & 0.040 & 0.723 & 0.227 & 0.837 & 0.113 & 0.554 & 0.396 & 0.673 & 0.277 \\ 
\midrule
Mean & 0.947 & 0.008 & 0.968 & 0.020 & 0.989 & 0.039 & 0.990 & 0.040 & 0.708 & 0.242 & 0.860 & 0.090 & 0.604 & 0.346 & 0.734 & 0.216 \\  
\bottomrule
\end{tabular}
\end{small} 
\end{table}

\newpage

\subsection*{Appendix~C: Canadian age-specific mortality rates}

\begin{figure}[!htb]
\centering
{\includegraphics[width=8.7cm]{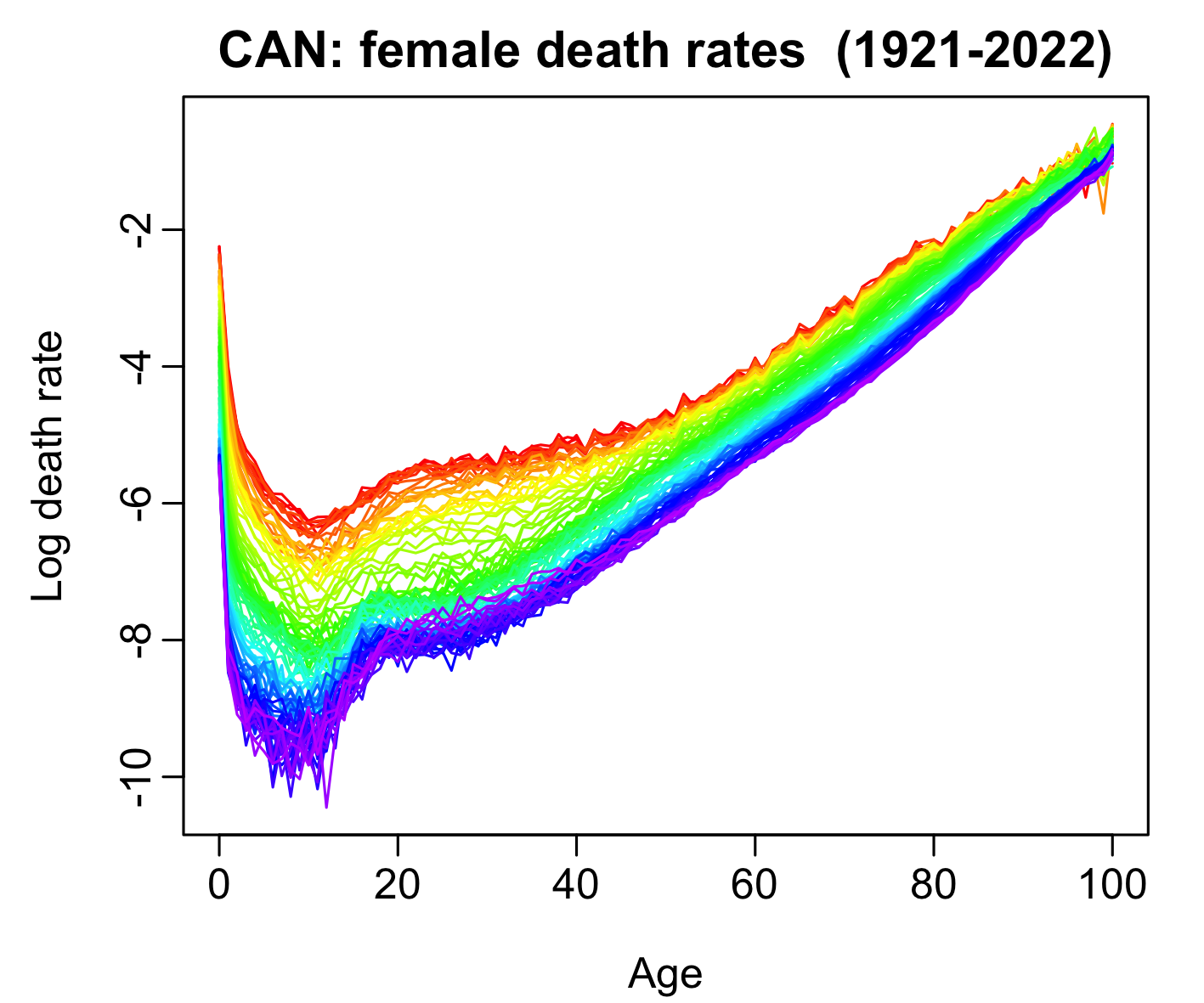}}
\quad
{\includegraphics[width=8.7cm]{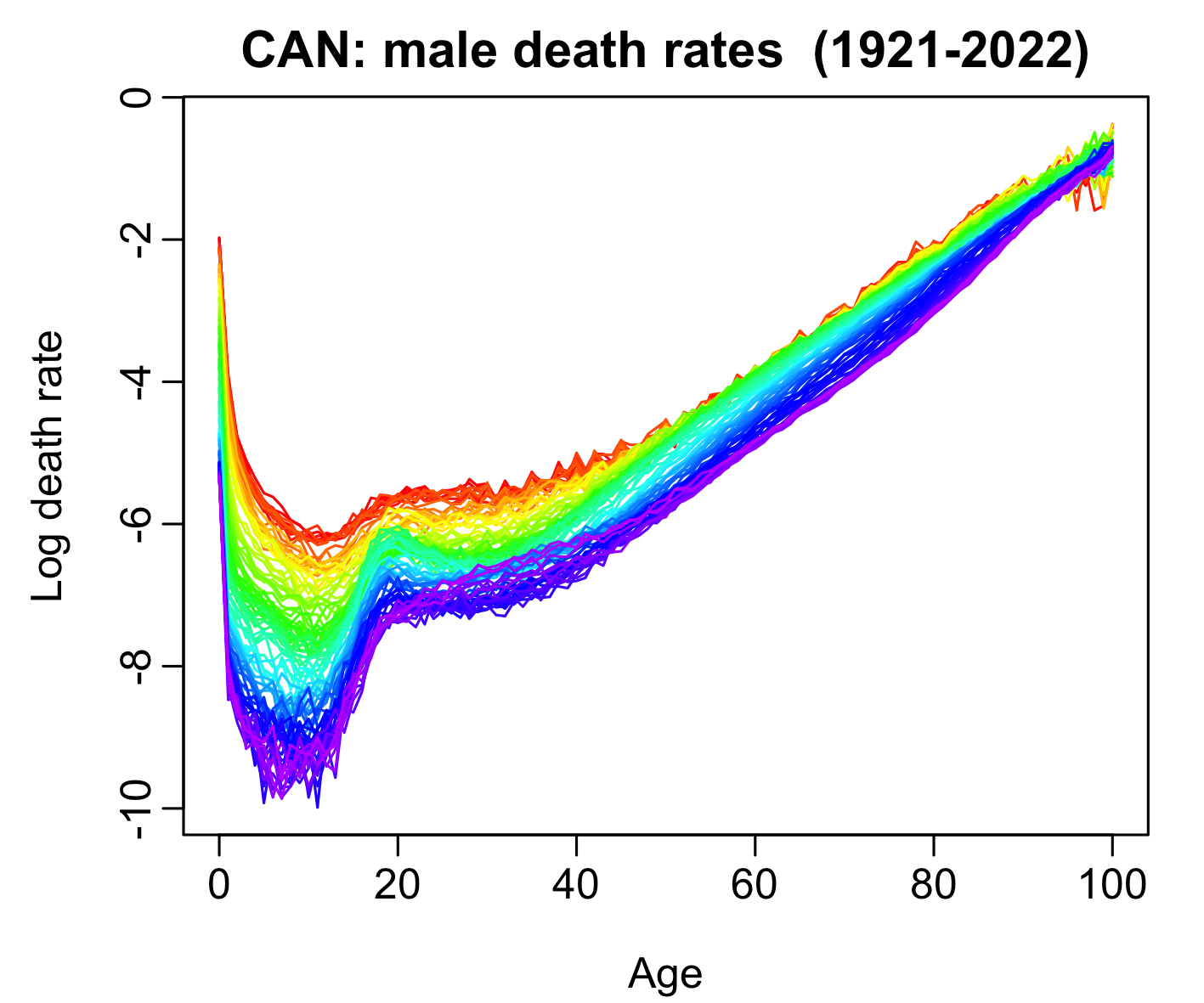}}
\\
{\includegraphics[width=8.7cm]{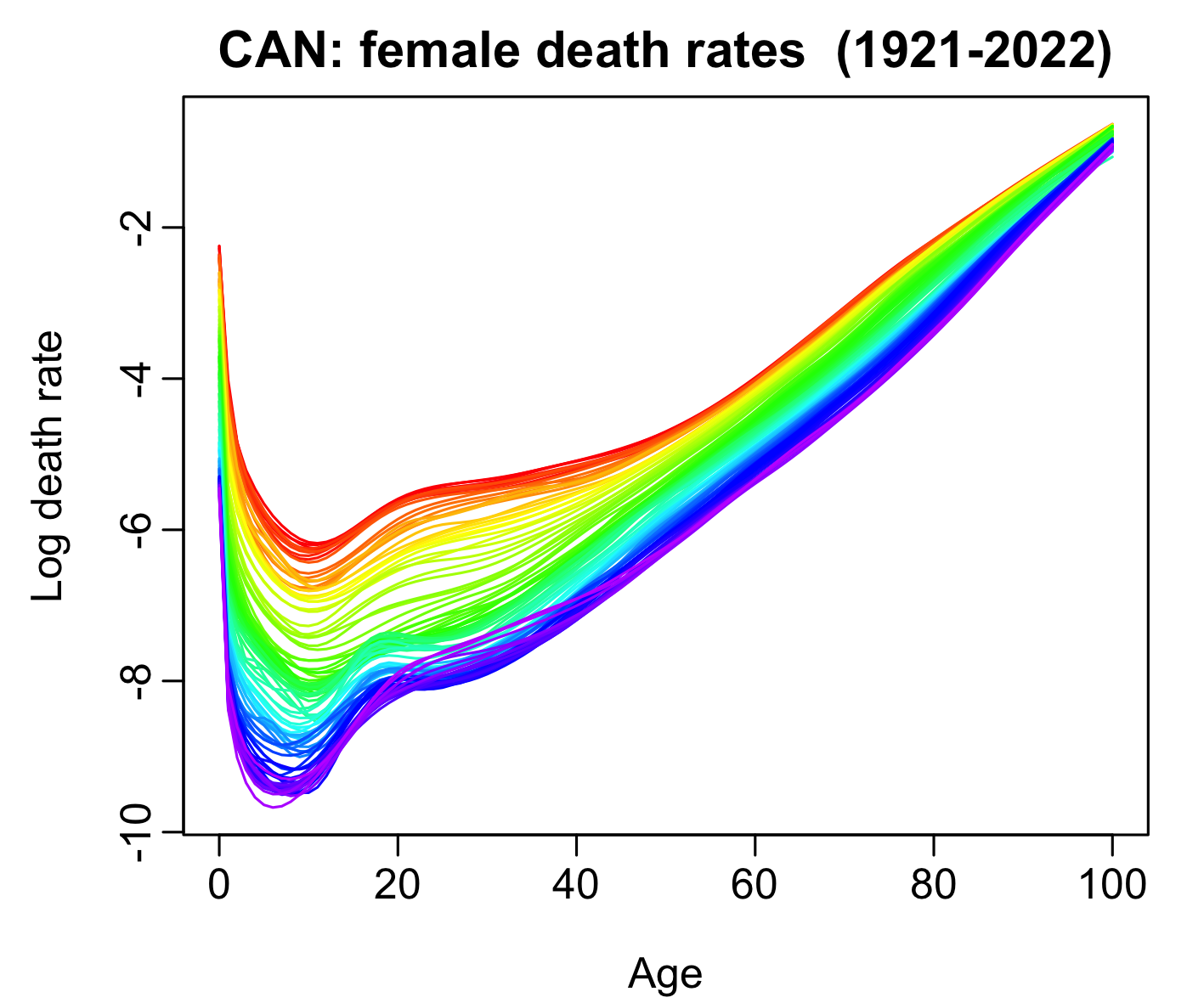}}
\quad
{\includegraphics[width=8.7cm]{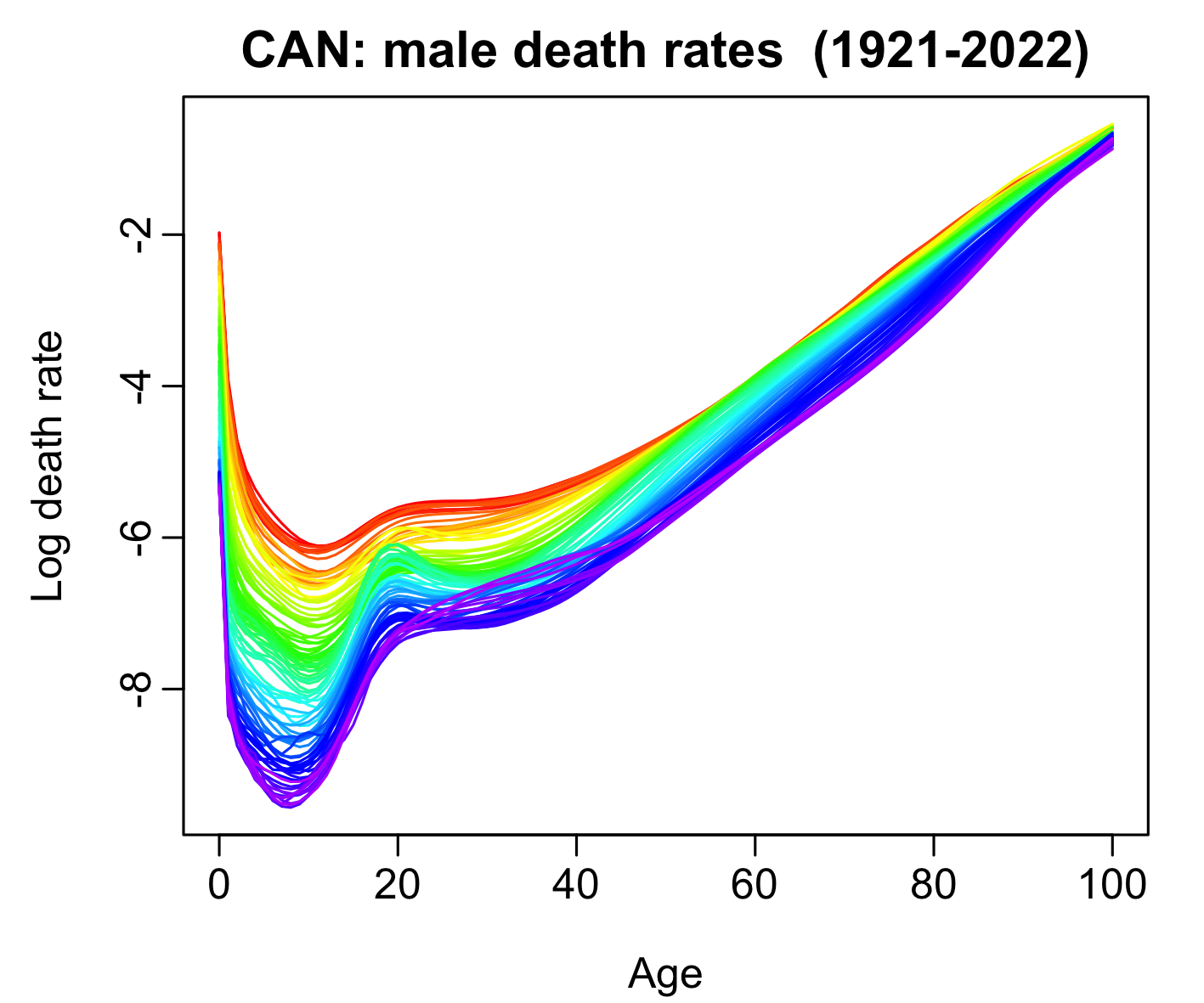}}
\caption{\small Rainbow plots of the original and smoothed age-specific mortality rates for the Canadian female and male data from 1921 to 2022.}\label{fig:A_3}
\end{figure}

We also analyze Canadian age- and sex-specific mortality rates spanning from 1921 to 2022, obtained from the \cite{HMD24}. Our study covers age groups from 0 to 99 in single years, with the final group covering ages 100 and above. We present rainbow plots for $\log$ mortality rates in Figure~\ref{fig:A_3}, where the data from the distant past are shown in red and the more recent data in purple.

In Table~\ref{tab:A_3}, we compute the horizon-specific empirical coverage probability and its coverage probability difference for the sd approach at the 80\% nominal coverage probability. Using data from 1921 to 1978, we computed the forecasts for the validation period from 1979 to 2000. For $h=1, 2,\dots,21$, we determine the optimal tuning parameters, with which we evaluate the empirical coverage probability based on the test period from 2001 to 2022. The sd approach achieves superior finite-sample coverage probability and often produces empirical coverage probability around the nominal level. 
\begin{table}[!htb]
\tabcolsep 0.04in
\renewcommand{\arraystretch}{0.98}
\centering
\caption{\small At the nominal coverage probability of 80\%, we compute the empirical coverage probability and its coverage probability difference between the sd approach and parametric approach for the Canadian data.}\label{tab:A_3}
\begin{small}
\begin{tabular}{@{}lrrrrrrrrrrrrrrrr@{}}
\toprule
& \multicolumn{8}{c}{sd approach}   & \multicolumn{8}{c}{parametric approach} \\
  \cmidrule(lr){2-9}  \cmidrule(lr){10-17}
& \multicolumn{4}{c}{Female} & \multicolumn{4}{c}{Male}  & \multicolumn{4}{c}{Female} & \multicolumn{4}{c}{Male} \\
\cmidrule(lr){2-5}\cmidrule(lr){6-9}\cmidrule(lr){10-13}\cmidrule(lr){14-17}	
& \multicolumn{2}{c}{Smooth} 	& \multicolumn{2}{c}{Raw}  & \multicolumn{2}{c}{Smooth} 	& \multicolumn{2}{c}{Raw} & \multicolumn{2}{c}{Smooth} 	& \multicolumn{2}{c}{Raw}  & \multicolumn{2}{c}{Smooth} 	& \multicolumn{2}{c}{Raw}  \\ 
\midrule
$h$ & ECP & CPD & ECP & CPD & ECP & CPD & ECP & CPD & ECP & CPD & ECP & CPD & ECP & CPD & ECP & CPD \\ 
\midrule
1 & 0.769 & 0.031 & 0.775 & 0.025 & 0.742 & 0.058 & 0.728 & 0.072 & 0.571 & 0.229 & 0.766 & 0.034 & 0.498 & 0.302 & 0.649 & 0.151 \\  
2 & 0.773 & 0.027 & 0.786 & 0.014 & 0.738 & 0.062 & 0.712 & 0.088 & 0.588 & 0.212 & 0.761 & 0.039 & 0.444 & 0.356 & 0.587 & 0.213 \\  
3 & 0.768 & 0.032 & 0.764 & 0.036 & 0.726 & 0.074 & 0.702 & 0.098 & 0.566 & 0.234 & 0.732 & 0.068 & 0.403 & 0.397 & 0.535 & 0.265 \\  
4 & 0.770 & 0.030 & 0.755 & 0.045 & 0.736 & 0.064 & 0.714 & 0.086 & 0.581 & 0.219 & 0.736 & 0.064 & 0.398 & 0.402 & 0.508 & 0.292 \\  
5 & 0.743 & 0.057 & 0.733 & 0.067 & 0.734 & 0.066 & 0.724 & 0.076 & 0.557 & 0.243 & 0.691 & 0.109 & 0.372 & 0.428 & 0.481 & 0.319 \\  
6 & 0.744 & 0.056 & 0.732 & 0.068 & 0.740 & 0.060 & 0.726 & 0.074 & 0.543 & 0.257 & 0.680 & 0.120 & 0.362 & 0.438 & 0.467 & 0.333 \\  
7 & 0.749 & 0.051 & 0.731 & 0.069 & 0.749 & 0.051 & 0.729 & 0.071 & 0.545 & 0.255 & 0.666 & 0.134 & 0.353 & 0.447 & 0.452 & 0.348 \\  
8 & 0.745 & 0.055 & 0.716 & 0.084 & 0.760 & 0.040 & 0.737 & 0.063 & 0.530 & 0.270 & 0.647 & 0.153 & 0.341 & 0.459 & 0.439 & 0.361 \\  
9 & 0.744 & 0.056 & 0.732 & 0.068 & 0.793 & 0.007 & 0.778 & 0.022 & 0.521 & 0.279 & 0.632 & 0.168 & 0.345 & 0.455 & 0.438 & 0.362 \\  
10 & 0.724 & 0.076 & 0.706 & 0.094 & 0.815 & 0.015 & 0.784 & 0.016 & 0.515 & 0.285 & 0.618 & 0.182 & 0.350 & 0.450 & 0.427 & 0.373 \\  
11 & 0.723 & 0.077 & 0.713 & 0.087 & 0.836 & 0.036 & 0.800 & 0.000 & 0.505 & 0.295 & 0.601 & 0.199 & 0.366 & 0.434 & 0.441 & 0.359 \\  
12 & 0.734 & 0.066 & 0.714 & 0.086 & 0.883 & 0.083 & 0.846 & 0.046 & 0.485 & 0.315 & 0.585 & 0.215 & 0.351 & 0.449 & 0.431 & 0.369 \\  
13 & 0.747 & 0.053 & 0.730 & 0.070 & 0.850 & 0.050 & 0.828 & 0.028 & 0.503 & 0.297 & 0.589 & 0.211 & 0.356 & 0.444 & 0.434 & 0.366 \\  
14 & 0.740 & 0.060 & 0.722 & 0.078 & 0.862 & 0.062 & 0.810 & 0.010 & 0.492 & 0.308 & 0.585 & 0.215 & 0.366 & 0.434 & 0.430 & 0.370 \\  
15 & 0.751 & 0.049 & 0.713 & 0.087 & 0.829 & 0.029 & 0.797 & 0.003 & 0.460 & 0.340 & 0.546 & 0.254 & 0.351 & 0.449 & 0.415 & 0.385 \\  
16 & 0.692 & 0.108 & 0.683 & 0.117 & 0.851 & 0.051 & 0.843 & 0.043 & 0.429 & 0.371 & 0.506 & 0.294 & 0.341 & 0.459 & 0.407 & 0.393 \\  
17 & 0.705 & 0.095 & 0.691 & 0.109 & 0.855 & 0.055 & 0.835 & 0.035 & 0.422 & 0.378 & 0.490 & 0.310 & 0.337 & 0.463 & 0.381 & 0.419 \\  
18 & 0.713 & 0.087 & 0.695 & 0.105 & 0.899 & 0.099 & 0.877 & 0.077 & 0.414 & 0.386 & 0.497 & 0.303 & 0.327 & 0.473 & 0.386 & 0.414 \\  
19 & 0.782 & 0.018 & 0.755 & 0.045 & 0.943 & 0.143 & 0.928 & 0.128 & 0.411 & 0.389 & 0.520 & 0.280 & 0.329 & 0.471 & 0.351 & 0.449 \\  
20 & 0.746 & 0.054 & 0.634 & 0.166 & 0.881 & 0.081 & 0.878 & 0.078 & 0.360 & 0.440 & 0.469 & 0.331 & 0.314 & 0.486 & 0.333 & 0.467 \\  
21 & 0.762 & 0.038 & 0.634 & 0.166 & 0.792 & 0.008 & 0.832 & 0.032 & 0.287 & 0.513 & 0.391 & 0.409 & 0.337 & 0.463 & 0.337 & 0.463 \\ 
\midrule
Mean & 0.744 & 0.056 & 0.720 & 0.080 & 0.810 & 0.057 & 0.791 & 0.055 & 0.490 & 0.310 & 0.605 & 0.195 & 0.364 & 0.436 & 0.444 & 0.356 \\ 
\bottomrule
\end{tabular}
\end{small} 
\end{table}

\newpage
In Table~\ref{tab:A_4}, we display the horizon-specific empirical coverage probability and its coverage probability difference for the sd approach at the 95\% nominal coverage probability. Compared with the parametric approach of \cite{HU07}, the sd approach achieves superior finite-sample coverage probability and often produces empirical coverage probability at and above the nominal level. 
\begin{table}[!htb]
\tabcolsep 0.04in
\renewcommand{\arraystretch}{0.9}
\centering
\caption{\small At the nominal coverage probability of 95\%, we compute the empirical coverage probability and its coverage probability difference between the sd approach and parametric approach.}\label{tab:A_4}
\begin{small}
\begin{tabular}{@{}lrrrrrrrrrrrrrrrr@{}}
\toprule
& \multicolumn{8}{c}{sd approach}   & \multicolumn{8}{c}{parametric approach} \\
  \cmidrule(lr){2-9}  \cmidrule(lr){10-17}
& \multicolumn{4}{c}{Female} & \multicolumn{4}{c}{Male}  & \multicolumn{4}{c}{Female} & \multicolumn{4}{c}{Male} \\
\cmidrule(lr){2-5}\cmidrule(lr){6-9}\cmidrule(lr){10-13}\cmidrule(lr){14-17}	
& \multicolumn{2}{c}{Smooth} 	& \multicolumn{2}{c}{Raw}  & \multicolumn{2}{c}{Smooth} 	& \multicolumn{2}{c}{Raw} & \multicolumn{2}{c}{Smooth} 	& \multicolumn{2}{c}{Raw}  & \multicolumn{2}{c}{Smooth} 	& \multicolumn{2}{c}{Raw}  \\ 
\midrule
$h$ & ECP & CPD & ECP & CPD & ECP & CPD & ECP & CPD & ECP & CPD & ECP & CPD & ECP & CPD & ECP & CPD \\ 
\midrule
1 & 0.918 & 0.032 & 0.907 & 0.043 & 0.892 & 0.058 & 0.881 & 0.069 & 0.738 & 0.212 & 0.902 & 0.048 & 0.683 & 0.267 & 0.845 & 0.105 \\  
2 & 0.925 & 0.025 & 0.915 & 0.035 & 0.876 & 0.074 & 0.864 & 0.086 & 0.746 & 0.204 & 0.897 & 0.053 & 0.613 & 0.337 & 0.778 & 0.172 \\  
3 & 0.906 & 0.044 & 0.900 & 0.050 & 0.855 & 0.095 & 0.851 & 0.099 & 0.733 & 0.217 & 0.865 & 0.085 & 0.560 & 0.390 & 0.710 & 0.240 \\  
4 & 0.916 & 0.034 & 0.903 & 0.047 & 0.855 & 0.095 & 0.858 & 0.092 & 0.735 & 0.215 & 0.865 & 0.085 & 0.538 & 0.412 & 0.670 & 0.280 \\  
5 & 0.895 & 0.055 & 0.881 & 0.069 & 0.864 & 0.086 & 0.848 & 0.102 & 0.704 & 0.246 & 0.837 & 0.113 & 0.528 & 0.422 & 0.642 & 0.308 \\  
6 & 0.878 & 0.072 & 0.867 & 0.083 & 0.871 & 0.079 & 0.858 & 0.092 & 0.694 & 0.256 & 0.822 & 0.128 & 0.497 & 0.453 & 0.615 & 0.335 \\  
7 & 0.877 & 0.073 & 0.860 & 0.090 & 0.876 & 0.074 & 0.858 & 0.092 & 0.691 & 0.259 & 0.806 & 0.144 & 0.494 & 0.456 & 0.612 & 0.338 \\  
8 & 0.869 & 0.081 & 0.844 & 0.106 & 0.894 & 0.056 & 0.878 & 0.072 & 0.673 & 0.277 & 0.788 & 0.162 & 0.498 & 0.452 & 0.601 & 0.349 \\  
9 & 0.866 & 0.084 & 0.854 & 0.096 & 0.895 & 0.055 & 0.875 & 0.075 & 0.656 & 0.294 & 0.778 & 0.172 & 0.487 & 0.463 & 0.588 & 0.362 \\  
10 & 0.847 & 0.103 & 0.840 & 0.110 & 0.902 & 0.048 & 0.882 & 0.068 & 0.648 & 0.302 & 0.750 & 0.200 & 0.489 & 0.461 & 0.586 & 0.364 \\  
11 & 0.851 & 0.099 & 0.847 & 0.103 & 0.920 & 0.030 & 0.903 & 0.047 & 0.634 & 0.316 & 0.737 & 0.213 & 0.506 & 0.444 & 0.593 & 0.357 \\  
12 & 0.869 & 0.081 & 0.829 & 0.121 & 0.954 & 0.004 & 0.938 & 0.012 & 0.628 & 0.322 & 0.714 & 0.236 & 0.515 & 0.435 & 0.598 & 0.352 \\  
13 & 0.869 & 0.081 & 0.847 & 0.103 & 0.935 & 0.015 & 0.940 & 0.010 & 0.630 & 0.320 & 0.727 & 0.223 & 0.525 & 0.425 & 0.592 & 0.358 \\  
14 & 0.834 & 0.116 & 0.826 & 0.124 & 0.964 & 0.014 & 0.978 & 0.028 & 0.617 & 0.333 & 0.722 & 0.228 & 0.528 & 0.422 & 0.601 & 0.349 \\  
15 & 0.838 & 0.112 & 0.832 & 0.118 & 0.944 & 0.006 & 0.978 & 0.028 & 0.597 & 0.353 & 0.712 & 0.238 & 0.519 & 0.431 & 0.580 & 0.370 \\  
16 & 0.825 & 0.125 & 0.818 & 0.132 & 0.963 & 0.013 & 0.972 & 0.022 & 0.562 & 0.388 & 0.673 & 0.277 & 0.506 & 0.444 & 0.567 & 0.383 \\  
17 & 0.810 & 0.140 & 0.784 & 0.166 & 0.972 & 0.022 & 0.979 & 0.029 & 0.558 & 0.392 & 0.649 & 0.301 & 0.493 & 0.457 & 0.569 & 0.381 \\  
18 & 0.808 & 0.142 & 0.810 & 0.140 & 0.980 & 0.030 & 0.974 & 0.024 & 0.562 & 0.388 & 0.657 & 0.293 & 0.497 & 0.453 & 0.543 & 0.407 \\  
19 & 0.879 & 0.071 & 0.901 & 0.049 & 0.998 & 0.048 & 0.978 & 0.028 & 0.564 & 0.386 & 0.653 & 0.297 & 0.460 & 0.490 & 0.532 & 0.418 \\  
20 & 0.911 & 0.039 & 0.908 & 0.042 & 0.990 & 0.040 & 0.970 & 0.020 & 0.521 & 0.429 & 0.640 & 0.310 & 0.426 & 0.524 & 0.475 & 0.475 \\  
21 & 0.881 & 0.069 & 0.876 & 0.074 & 0.980 & 0.030 & 0.941 & 0.009 & 0.480 & 0.470 & 0.634 & 0.316 & 0.431 & 0.519 & 0.480 & 0.470 \\  
\midrule
Mean & 0.870 & 0.080 & 0.859 & 0.091 & 0.923 & 0.046 & 0.914 & 0.053 & 0.637 & 0.313 & 0.754 & 0.196 & 0.514 & 0.436 & 0.608 & 0.342 \\  
\bottomrule
\end{tabular}
\end{small} 
\end{table}
  
\newpage
\bibliographystyle{agsm}
\bibliography{PI_LTDC}

\end{document}